\documentclass[10pt,twocolumn,british]{elsarticle}
\usepackage[latin9]{inputenc}
\usepackage{color}
\usepackage{amsmath}
\usepackage{graphicx}

\makeatletter

\providecommand{\tabularnewline}{\\}

\usepackage{a4wide}

\usepackage[hypertex]{hyperref}

\usepackage[T1]{fontenc}
\usepackage{ae,aecompl}

\usepackage{framed}
\usepackage{amsmath}
\usepackage{bm}% for bold math symbols such as sigma.
\usepackage{verbatim}

\usepackage{ifthen}
\newboolean{includeNotes}
\setboolean{includeNotes}{false}
\newcommand{\T}{Table~}

\newcommand{\F}{Figure~}

\newcommand{\Eq}{Eq.~}
\newcommand{\Eqs}{Eqs.~}

\renewcommand{\S}{\textsection}

\makeatother

\usepackage{babel}

\begin{document}

\title{Maximum wave-power absorption by attenuating line absorbers under
volume constraints}

\author{Paul~Stansell\corref{cor1}}
\ead{p.stansell@pelamiswave.com}

\author{David J.~Pizer}
\ead{d.pizer@pelamiswave.com}

\cortext[cor1]{Corresponding author}

\address{Pelamis Wave Power Ltd., 31 Bath Road, Edinburgh, UK, EH6 7AH}
\begin{abstract}
This work investigates the consequences of imposing a volume constraint
on the maximum power that can be absorbed from progressive regular
incident waves by an attenuating line absorber heaving in a travelling
wave mode. Under assumptions of linear theory an equation for the
maximum absorbed power is derived in terms of two dimensionless independent
variables representing the length and the half-swept volume of the
line absorber. The equation gives the well-known result for a point
absorber wave energy converter in the limit of zero length and it
gives Budal's upper bound in the limit of zero volume. The equation
shows that the maximum power absorbed by a heaving point absorber
is limited regardless of its volume, while for a heaving line absorber
whose length tends to infinity the maximum power is proportional to
its swept volume, with no limit. Power limits arise for line absorbers
of practical lengths and volumes but they can be multiples of those
achieved for point absorbers of similar volumes. This conclusion has
profound implications for the scaling and economics of wave energy
converters.\end{abstract}
\maketitle
\clearpage

\section{Introduction }

A large variety of wave energy converters (WECs) have been proposed,
with a few demonstrated at full scale. This paper is concerned with
the following two categories of buoyancy-driven heaving WECs: the
archetypal point absorber which is a single semi-submerged float with
horizontal dimensions $a$ that are small compared to the wavelength
$\lambda$ of incident wave; and the archetypal attenuating line absorber
which is a continuous line of semi-submerged floats with its long
dimension orientated in the direction of travel of the incident wave,
its width $a$ small compared to the wavelength and its length $l$
of the order of a wavelength.

It is well established from linear water-wave theory that the upper
limit of the capture width of a heaving point absorber is $\lambda/2\pi$
(see, for example, \citet{Evans1976,Mei1976,Newman1976}). As $\lambda$
increases, however, reaching a capture width of $\lambda/2\pi$ requires
an increasingly large swept volume. Of more practical significance
are the results of \citet{Evans1981} and \citet{Pizer93} that account
for a constraint on the motion---and, therefore, the swept volume---of
the float.

\citet{Farley82} provides a theory for the capture width of an attenuating
line absorber oscillating in a travelling wave mode, but he does not
incorporate a motion constraint. \citet{Newman1979} does consider
a motion constraint but his theory is restricted to modes of motion
that have uniform temporal phase and so he does not solve explicitly
for the more practical travelling wave mode. The present work follows
the approaches of \citet{Farley82} and \citet{Rainey01} in considering
the travelling wave mode response, and it derives the maximum capture
width with a motion constraint to limit the maximum swept volume of
the device.

The method calculates the capture width by considering the radiated
and diffracted waves from a line composed of a continuum of point
wave sources and their interaction with the incident wave in the far-field.
The motion constraint is applied by matching the far-field radiated
wave amplitude with the swept volume of a line segment using the relative
motion hypothesis. Example calculations are provided for point and
line absorbers of different volumes in wave conditions of practical
interest. The effects of variations in wavelength and wave height
are also illustrated.

\section{Background Theory \label{sec:Theory}}

\subsection{Coordinate systems}

The coordinate systems and related notation are described here. Let
$\Sigma$ denote an inertial reference frame described by the Cartesian
coordinates $(x,y,z)\equiv(x_{1},x_{2},x_{3})$ with basis $\left(\hat{\mathbf{e}}_{x},\hat{\mathbf{e}}_{y},\hat{\mathbf{e}}_{z}\right)\equiv\left(\hat{\mathbf{e}}_{1},\hat{\mathbf{e}}_{2},\hat{\mathbf{e}}_{3}\right)$
and by the cylindrical polar coordinates $(r,\theta,z)$ with basis
$\left(\hat{\mathbf{e}}_{r},\hat{\mathbf{e}}_{\theta},\hat{\mathbf{e}}_{z}\right)$.
The origins of these coordinate systems coincide, the $z$-axis is
directed vertically upward, the plane of the undisturbed free surface
is at $z=0$ and the fluid bottom is at $z=-h$. When $\theta=0$
the unit vectors $\hat{\mathbf{e}}_{r}$ and $\hat{\mathbf{e}}_{x}$
are equal. For any vector $\mathbf{n}$, the following notations are
equivalent: $\mathbf{n}\cdot\boldsymbol{\nabla}\equiv n_{\alpha}\,\partial/\partial x_{\alpha}\equiv\partial/\partial n\equiv\partial_{n}$,
where $\boldsymbol{\nabla}$ is the gradient operator and $\alpha$
can take values from $\{1,2,3\}$ or $\{x,y,z\}$.

\subsection{Governing equations \label{sec:Governing-equations}}

Consider a rigid body floating on the surface of an inviscid, irrotational,
incompressible fluid and interacting with a plane incident wave. All
motions are assumed to be time-harmonic with angular frequency $\omega$.
It is assumed that the body-induced perturbations are small so that
the linearised theory of the interaction of water waves and structures
can be applied and higher-order effects can be neglected.

Under these assumptions the fluid is described by either its real
velocity potential, $\Phi$, or by the time-independent complex amplitude
of the velocity potential, $\phi$, where \begin{equation}
\Phi=\Re\left\{ \phi\,\mathrm{e}^{-i\omega t}\right\} .\label{eq:Phi}\end{equation}
The velocity potential, $\phi$, must satisfy the Laplace equation,
\begin{equation}
\nabla^{2}\phi=0,\quad-h<z<0.\label{eq:Laplace_eq}\end{equation}
In the linearised approximation for small amplitude waves and constant
atmospheric pressure at $z=0$, the boundary conditions at the sea
bed and the free surface are\begin{eqnarray}
\frac{\partial\phi}{\partial z} & = & 0,\quad z=-h,\label{eq:BC_-h}\\
-\omega^{2}\phi+g\frac{\partial\phi}{\partial z} & = & 0,\quad z=0,\label{eq:BC_z0}\end{eqnarray}
where $g$ is the acceleration due to gravity.

The boundary condition on the wetted surface, denoted by $S_{B}$,
of a heaving body is obtained by equating the normal velocity of the
body to the normal velocity of the fluid on $S_{B}$. The result is\begin{equation}
\mathbf{n}\cdot\boldsymbol{\nabla}\phi=-i\omega\, n_{z}A_{z}\quad\mbox{on }S_{B},\label{eq:BC_floating_body}\end{equation}
where $\mathbf{n}$ is the unit vector in the direction of the normal
pointing into the body at any point on $S_{B}$, $n_{z}$ is the component
of $\mathbf{n}$ in the direction of $\hat{\mathbf{e}}_{z}$, and
$A_{z}$ is the complex amplitude of the oscillatory heave motion
of the body about its mean position.

\subsection{Extraction of wave energy \label{sub:Extraction-of-wave}}

The time averaged power, denoted by $P$, transferred from the fluid
to the body is obtained by integrating the time average of the fluid
pressure, $p$, times the fluid velocity, $\partial\Phi/\partial n$,
over $S_{B}$. This may be expressed as%
\footnote{Within the linear theory applied here, it is acceptable to integrate
over the time average of the position of the surface of the body as
integrating over the instantaneous position of the surface of the
body adds at most a second-order term to the time average of the absorbed
power.%
} \begin{equation}
P=\iint_{S_{B}}\left(\frac{1}{T}\int_{0}^{T}p\,\frac{\partial\Phi}{\partial n}\, dt\right)\, dS,\label{eq:W60_8.3}\end{equation}
where $T=2\pi/\omega$ is the period of oscillation. Note that $P$
is the time average of the rate of change of work done on the body
so that a positive value of $P$ represents energy absorbed by the
body from the fluid.

The pressure field in the fluid is given by the linearised Bernoulli
equation which can be written as\begin{equation}
\frac{\partial\Phi}{\partial t}+gz+\frac{p}{\rho}=\frac{p_{0}}{\rho},\label{eq:Bernoulli}\end{equation}
where $p_{0}$ is the pressure at $z=0$, $g$ is the acceleration
due to gravity and $\rho$ is the fluid density. In the absence of
a wave it is assumed that $p=p_{0}$ is constant at $z=0$.

Substituting $\Phi$ from (\ref{eq:Phi}) and $p$ from (\ref{eq:Bernoulli})
into the integrand in (\ref{eq:W60_8.3}), it can be shown that\begin{equation}
\frac{1}{T}\int_{0}^{T}p\,\frac{\partial\Phi}{\partial n}\, dt=\frac{i\omega\rho}{4}\left(\phi\,\frac{\partial\bar{\phi}}{\partial n}-\bar{\phi}\,\frac{\partial\phi}{\partial n}\right),\label{eq:time_average}\end{equation}
where $\bar{\phi}$ is the complex conjugate of $\phi$ and use has
been made of  the fact that the integrals of the oscillatory terms
are zero. Substituting (\ref{eq:time_average}) into (\ref{eq:W60_8.3})
gives\begin{equation}
P=\frac{i\omega\rho}{4}\iint_{S_{B}}\left(\phi\,\frac{\partial\bar{\phi}}{\partial n}-\bar{\phi}\,\frac{\partial\phi}{\partial n}\right)\, dS.\label{eq:dE/dt_Sb}\end{equation}

Application of Green's second identity   to the integral in (\ref{eq:dE/dt_Sb}),
and using the homogeneous boundary conditions (\ref{eq:BC_-h}) and
(\ref{eq:BC_z0}), gives the integral over the body's surface in terms
of an integral over a far-field control surface. The result is\begin{multline}
\iint_{S_{B}}\left(\phi\,\frac{\partial\bar{\phi}}{\partial n}-\bar{\phi}\,\frac{\partial\phi}{\partial n}\right)\, dS\\
+\iint_{S_{C}}\left(\phi\,\frac{\partial\bar{\phi}}{\partial n}-\bar{\phi}\,\frac{\partial\phi}{\partial n}\right)\, dS=0,\label{eq:Greens2nd}\end{multline}
where $S_{C}$ is a cylindrical control surface whose axis is in the
direction $\hat{\mathbf{e}}_{z}$ and which encircles the body. Substituting
(\ref{eq:Greens2nd}) into (\ref{eq:dE/dt_Sb}) converts the integral
over the body's surface into an integral over the far-field control
surface, and it follows that\begin{equation}
P=-\frac{i\omega\rho}{4}\iint_{S_{C}}\left(\phi\,\frac{\partial\bar{\phi}}{\partial n}-\bar{\phi}\,\frac{\partial\phi}{\partial n}\right)\, dS.\label{eq:dE/dt_Sc}\end{equation}
Forthcoming algebra is simplified by substituting\[
\phi\,\frac{\partial\bar{\phi}}{\partial n}-\bar{\phi}\,\frac{\partial\phi}{\partial n}\equiv2i\Im\left\{ \phi\,\frac{\partial\bar{\phi}}{\partial n}\right\} \]
into (\ref{eq:dE/dt_Sc}) to give%
\footnote{Note that (\ref{eq:dE/dt_Sc}) and (\ref{eq:dE/dt_Sc_2}) appear as
unnumbered equations at the top of page 320 in \citet{Mei89}.%
} \begin{equation}
P=\frac{\omega\rho}{2}\Im\left\{ \iint_{S_{C}}\phi\,\frac{\partial\bar{\phi}}{\partial n}\, dS\right\} .\label{eq:dE/dt_Sc_2}\end{equation}
This equation expresses the power, $P$, transferred from the fluid
to the body in terms of the total power entering the control volume
bounded by the control surface, $S_{C}$.

\subsection{Linear decomposition of velocity potential}

According to the linearised theory of the interaction of water waves
with heaving bodies, the total velocity potential of the fluid may
be approximated by\begin{equation}
\phi=A_{0}\,\varphi_{0}+A_{0}\,\varphi_{d}+A_{z}\,\varphi_{z},\label{eq:phi_comps}\end{equation}
where: $\varphi_{0}$ and $\varphi_{d}$ are the velocity potentials
per unit complex amplitude for the incident and diffracted waves respectively;
$A_{0}$ is the complex amplitude of the incident wave; $\varphi_{z}$
is the velocity potential per unit complex amplitude of the oscillating
heave motion of the rigid-body; and $A_{z}$ is the complex amplitude
of the oscillating heave-mode of the body as introduced previously
in (\ref{eq:BC_floating_body}). Each of the velocity potentials $\varphi_{0}$
, $\varphi_{d}$ and $\varphi_{z}$ satisfy the Laplace equation (\ref{eq:Laplace_eq}),
the sea bed condition (\ref{eq:BC_-h}), and the free-surface condition
(\ref{eq:BC_z0}). Furthermore, $\varphi_{d}$ and $\varphi_{z}$
satisfy the Sommerfield radiation condition at large distances from
the body. 

In this treatment the incident wave is a plane wave travelling in
the positive $x$-direction with a velocity potential given by\begin{equation}
\phi_{0}=A_{0}\varphi_{0}=-iA_{0}\frac{g}{\omega}\,\frac{\cosh(k(z+h))}{\cosh(kh)}\,\mathrm{e}^{ikx},\label{eq:varphi_0}\end{equation}
and a dispersion relationship given by\begin{equation}
\omega^{2}=gk\,\tanh(kh),\label{eq:dispersion}\end{equation}
where $\omega$ is the wave's angular frequency and $k$ is its wavenumber. 

The linearised boundary conditions on $S_{B}$ are obtained by substituting
$\phi$ from (\ref{eq:phi_comps}) into the boundary condition (\ref{eq:BC_floating_body}).
Setting $A_{z}=0$ gives\begin{equation}
\mathbf{n}\cdot\boldsymbol{\nabla}\varphi_{d}=-\mathbf{n}\cdot\boldsymbol{\nabla}\varphi_{0}\quad\mbox{on }S_{B},\label{eq:rel_mot_bc1}\end{equation}
whilst setting $A_{0}=0$ gives\begin{equation}
\mathbf{n}\cdot\boldsymbol{\nabla}\varphi_{z}=-i\omega\, n_{z}\quad\mbox{on }S_{B}.\label{eq:rel_mot_bc2}\end{equation}

\subsection{Relative motion hypothesis}

This section summarises the application of the relative motion hypothesis
and matched asymptotic expansions used by \citet{McIver1994} to give
the leading order relationship between the near and far-field expressions
of $\varphi_{d}$ and $\varphi_{z}$. Note that the finite-depth analysis
of \citeauthor{McIver1994} \citep[\S9 and \S10]{McIver1994} is not
applicable to the problem here as it applies to cases in which the
body dimensions are of the same order as the water depth. We therefore
proceed with the finite-depth generalisations of the \citeauthor{McIver1994}'s
infinite depth results \citep[\S7]{McIver1994}.

It is assumed that $S_{B}$ is centred at the origin of the coordinate
system and that its extent is small compared with the wavelength of
the incident wave. Under these assumptions it is consistent with the
analysis presented here to approximate $\varphi_{0}$ on $S_{B}$
by using just the terms up to first-order in $ka$ in the Taylor expansion
of $\varphi_{0}$ from (\ref{eq:varphi_0}) with respect to the dimensionless
variables $kx$ and $kz$ about the origin $x=y=0$. That is,\begin{equation}
\varphi_{0}=-i\frac{g}{\omega}\left(1+k\left(ix+z\,\tanh(kh)\right)+O(ka)^{2}\right).\label{eq:taylor_phi0}\end{equation}
This is the finite-depth equivalent to (158) in \citep{McIver1994}.
Equation~(\ref{eq:taylor_phi0}) can be substituted into the body
boundary conditions (\ref{eq:rel_mot_bc1}), (\ref{eq:rel_mot_bc2}),
and the equivalent surge boundary condition for $\varphi_{x}$, to
show that in the vicinity of the body $\varphi_{d}$ may be expressed
as a linear combination of $\varphi_{x}$ and $\varphi_{z}$, that
is,\[
\varphi_{d}\sim-i\,\frac{\varphi_{x}}{\tanh(kh)}-\varphi_{z}\quad\mbox{on }S_{B}.\]
This is the finite depth equivalent to the near-field relationship given
by (160) in \citep{McIver1994}.

The far-field relationship between the diffraction and radiation potentials
is obtained by \citeauthor{McIver1994} \citep[\S7]{McIver1994} using
matched asymptotic analysis. He shows that for an axisymmetric surface-piercing
body the far-field velocity potential of radiated waves arising from
heave motions is proportional to $(ka)^{2}$, whereas that from surge
motions is proportional to $(ka)^{3}$. In particular, using \citeauthor{McIver1994}'s
\citep{McIver1994} far-field expressions (132) and (161), it is clear
that the leading order terms in the far-field diffracted and radiated
waves are related by $\varphi_{d}\sim-\varphi_{z}$. Thus, when performing
the integral over the far-field control surface $S_{C}$ in (\ref{eq:dE/dt_Sc_2})
it is consistent with the approximations used here to write\begin{equation}
\varphi_{d}\sim-\varphi_{z}\quad\mbox{on }S_{C}.\label{eq:rel_mot_hyp_far_field}\end{equation}

\section{Heaving attenuating line absorber \label{sec:attenuator}}

\subsection{Capture width for unlimited volume}

The slender body approximation is used to derive an expression for
the capture width of an attenuating line absorber, as in \citet{Newman1979}.
This approximation gives an expression for the far-field radiated
velocity potential at a distance $r$ from a slender body. For a body
of length $l$ and beam $b$ it states that, provided $b\ll\lambda$
and $l\ll r$, an approximation for the far-field radiated velocity
potential at $r$ can be obtained by integrating over the contributions
from the infinitesimal elements of waterplane area of the body, $b\,\delta l$
where $\delta l\ll\lambda$. Denoting by $\delta\phi_{n,z}$ the velocity
potential from the heave motions of the $n$th infinitesimal element,
it follows that the far-field velocity potential due to the heave
motions of all the infinitesimal elements comprising the slender body
is given by\begin{equation}
\phi_{z}=\int\delta\phi_{n,z}.\label{eq:slender_body_main}\end{equation}

To express each $\delta\varphi_{n,z}$, consider the only source of
waves in a fluid to be $A_{z}$ amplitude heave oscillations of the
$n$th infinitesimal element of the body. Let $(r'_{n},\theta'_{n},z'$)
be the polar coordinates in the coordinate system for which the infinitesimal
element is centred at $r'_{n}=0$. The velocity potential per unit
complex amplitude of this infinitesimal element at a large distance
$r'_{n}$ from the centre of the body is given by%
\footnote{See, for example, \citet{McIver1994}, in particular see \Eqs(10)
, (143) and the unnumbered equation immediately following (143) which
is the Kochin function, denoted by $\mathcal{A}_{3}$ in \citeauthor{McIver1994}'s
notation.%
}\begin{multline}
\delta\varphi_{n,z}(r'_{n},\theta'_{n},z')\sim-\frac{g}{\omega}\,\delta H_{z}(\theta'_{n})\left(\frac{2}{\pi kr'_{n}}\right)^{1/2}\!\!\times\\
\frac{\cosh(k(z'+h))}{\cosh(kh)}\mathrm{e}^{ikr'_{n}-i\pi/4},\quad kr'_{n}\to\infty,\label{eq:varphi_r'}\end{multline}
where $\delta H_{z}(\theta'_{n})$ is the Kochin function \citep[for a definition, see][]{Mei89},
which is a dimensionless function describing the angular dependency
of the radiated wave amplitude of the heave motion in the limit $kr'_{n}\to\infty$.
Generally, for a heaving body with infinitesimal waterplane area $\delta S_{w}$,
the Kochin function is independent of $\theta$ and is given by $\delta H_{z}=\frac{1}{2}k^{2}\,\delta S_{w}$
\citep[see $\mathcal{A}_{3}$  on page~22 of][]{McIver1994}. Thus,
for the infinitesimal rigid elements comprising a slender body of
constant beam, $\delta S_{w}=b\,\delta l$, it follows that\begin{equation}
\delta H_{z}=\frac{1}{2}k^{2}b\,\delta l.\label{eq:H_3b}\end{equation}
Here, the assumption is made that the waterplane area of the element
is not a function of the depth of submergence of that element. This
assumption is strictly valid either for infinitesimal relative motions
or for vertical wall-sided elements or both. 

To apply the slender body approximation in (\ref{eq:slender_body_main}),
let the coordinates in $\Sigma$ of the centre of the $n$th infinitesimal
rigid-body element be $x=x_{n}$ along the line $z=y=0$ such that
$-l/2\leq x_{n}\leq l/2$ and let the length of the element be $\delta l=\delta x_{n}$.
Denote by\begin{equation}
\delta\phi_{n,z}(x,y,z;x_{n})=A_{n,z}\,\delta\varphi_{n,z}(x,y,z;x_{n})\label{eq:delta-phi_n,a}\end{equation}
the far-field velocity potential at $(x,y,z)$ in $\Sigma$ due to
heave oscillations of the infinitesimal source at $x=x_{n}$. Let
$\Sigma'_{n}$ denote the coordinate system with the same Cartesian
basis vectors as $\Sigma$ but with its origin shifted to $(x{}_{n},0,0)$
as measured from $\Sigma$. Thus, in $\Sigma'_{n}$ the $n$th infinitesimal
element is at $(x'_{n},y'_{n},z'_{n})=(0,0,0)$ in Cartesian coordinates
or, equivalently, at $r'_{n}=0$ in polar coordinates $(r'_{n},\theta'_{n},z'_{n})$.
The Cartesian coordinate transformations between $\Sigma'_{n}$ and
$\Sigma$ are\begin{eqnarray}
x'_{n} & = & x-x_{n},\label{eq:x'n}\\
y'_{n} & = & y,\label{eq:y'n}\\
z'_{n} & = & z.\label{eq:z'n}\end{eqnarray}
Substituting (\ref{eq:x'n})--(\ref{eq:z'n}) into $r'_{n}=\left(x_{n}'^{2}+y_{n}'^{2}\right)^{1/2}$,
expanding, and using $r^{2}=x^{2}+y^{2}$ and $x=r\,\cos\theta$,
it can be shown that \begin{equation}
r'_{n}=r-x_{n}\cos\theta+O\left(\frac{x_{n}^{2}}{r}\right).\label{eq:r'_n}\end{equation}

\begin{figure}
\noindent \begin{centering}
\includegraphics[width=0.5\textwidth]{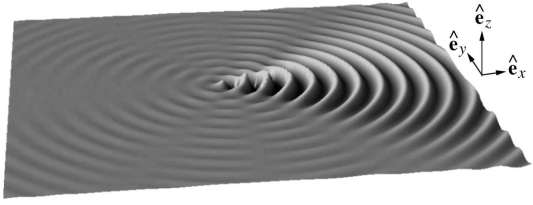}
\par\end{centering}

\caption{\label{fig:schematic_eta}Schematic of the surface elevation of the
wave radiated from a line of heaving elements oscillating on otherwise
still water. The line has a length of $l/\lambda=2$ and is oscillating
in a travelling wave mode with velocity potential given by (\ref{eq:phi_alpha_bess}).}
\ifthenelse{\boolean{includeNotes}}{

\textcolor{green}{Figure obtained from Dave. Original file was emailed
to me by Dave as a colour jpeg file called 11mono\_pel200m.jpg at
Tue Dec 6 18:07:04 2011. I used gimp to change the background to white,
and write it out as a black and white tiff file. (AOR accepts tiff
files, but not jpeg files).}

}

\end{figure}

Substituting (\ref{eq:H_3b}), (\ref{eq:r'_n}) and $\delta l=\delta x_{n}$
into (\ref{eq:varphi_r'}) and neglecting terms of order $O(x_{n}^{2}/r)$
and higher, $\delta\varphi_{n,z}(r'_{n},\theta'_{n},z)$ is transformed
from the $\Sigma'_{n}$ reference frame to the $\Sigma$ reference
frame yielding\begin{multline}
\delta\varphi_{n,z}(r,\theta,z;x_{n})\sim-\frac{gk^{2}b\,\delta x_{n}}{\omega\left(2\pi kr\right)^{1/2}}\times\\
\frac{\cosh(k(z+h))}{\cosh(kh)}\mathrm{e}^{ikr-i\pi/4}\mathrm{e}^{-ikx_{n}\cos\theta},\\
\quad kr\to\infty.\label{eq:delta_varphi_r}\end{multline}
For an attenuating line absorber composed of heaving elements oscillating
with relative phases that give the dynamics of a travelling wave moving
in the positive $x$-direction along the length of the attenuator,
the complex amplitude can be written as the function of $x_{n}$ given
by\begin{equation}
A_{n,z}(x_{n})=A_{z}\,\mathrm{e}^{ikx_{n}},\label{eq:A_n,x}\end{equation}
where $A_{z}$ is the complex amplitude of the heave-mode oscillation
of the element located at the origin of the $\Sigma$ reference frame.
Substituting (\ref{eq:A_n,x}) into (\ref{eq:delta-phi_n,a}), it
follows that\begin{equation}
\delta\phi_{n,z}(r,\theta,z;x_{n})=A_{z}\,\mathrm{e}^{ikx_{n}}\,\delta\varphi_{n,z}(r,\theta,z;x_{n}).\label{eq:delta_phi_r}\end{equation}
From (\ref{eq:slender_body_main}), (\ref{eq:delta_varphi_r}) and
(\ref{eq:delta_phi_r}), $\phi_{z}(r,\theta,z)$ can be evaluated
 as\begin{multline}
\phi_{z}(r,\theta,z)\sim-\frac{gk^{2}A_{z}\, bl}{\omega\left(2\pi kr\right)^{1/2}}\frac{\cosh(k(z+h))}{\cosh(kh)}\times\\
j_{0}\negmedspace\left(\frac{kl}{2}\left(1-\cos\theta\right)\right)\,\mathrm{e}^{ikr-i\pi/4},\quad kr\to\infty,\label{eq:phi_alpha_bess}\end{multline}
where $j_{0}$ is the zeroth-order spherical Bessel function of the
first kind (also known as the \emph{sine cardinal} or \emph{sinc}
function). All of the $\theta$-dependency in (\ref{eq:phi_alpha_bess})
is within the spherical Bessel function. The heave-mode oscillations
have wavelength $\lambda$ and are travelling in the $\hat{\mathbf{e}}_{x}$-direction
along the length, $l$, of the line absorber. Figure~\ref{fig:schematic_eta}
shows a schematic of the $\theta$-dependency of the surface elevation
of the radiated wave, given by $\eta=-g^{-1}\left.\partial\phi_{z}/\partial t\right|_{z=0}$,
for $\phi_{z}$ in (\ref{eq:phi_alpha_bess}) with $l/\lambda=2$.
The concentration of waves in the $\hat{\mathbf{e}}_{x}$-direction
resulting from these phased motions is clearly visible in the figure. 

\begin{figure}
\noindent \begin{centering}
\includegraphics[width=0.5\textwidth]{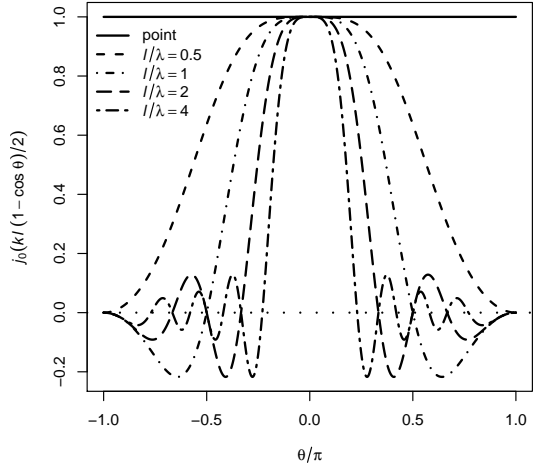}
\par\end{centering}

\caption{\label{fig:j_0}Plots of the function $j_{0}\negmedspace\left(\frac{kl}{2}\left(1-\cos\theta\right)\right)$.
This function describes the $\theta$-dependency of the surface elevation
of the radiated waves and that of the velocity potential, $\phi_{z}$,
from (\ref{eq:phi_alpha_bess}). The $\theta$-dependency plots for
a point absorber and four line absorbers of different lengths $l/\lambda$
are shown. }
\ifthenelse{\boolean{includeNotes}}{

\textcolor{green}{Figures produced by \textasciitilde{}/opd/documents/line\_absorbers/R-analysis/power\_plots\_AOR.R.}

}

\end{figure}

For a given amplitude $A_{z}$ and waterplane area $bl$, and since
$j_{0}(0)=1$, it can be seen from (\ref{eq:phi_alpha_bess}) that
in the limit $kr\to\infty$ the ratio of $\phi_{z}$ for a line absorber
of length $l$ to that of a point absorber is simply $j_{0}\negmedspace\left(\frac{kl}{2}\left(1-\cos\theta\right)\right)$.
Since the surface elevation of the radiated waves is proportional
to $\phi_{z}$ at $z=0$, the ratio of the surface elevation of a
line absorber to that of a point absorber is also $j_{0}\negmedspace\left(\frac{kl}{2}\left(1-\cos\theta\right)\right)$.
Graphs of $j_{0}\negmedspace\left(\frac{kl}{2}\left(1-\cos\theta\right)\right)$
for a point absorber and four line absorbers of different lengths
are shown in \F\ref{fig:j_0}. From this figure it is clear that
the point absorber radiates waves in all directions with the same
amplitude, whereas the line absorbers radiate at the same amplitude
as the point absorber in the $\theta=0$ or $\hat{\mathbf{e}}_{x}$-direction,
but they radiate at reduced amplitudes in other directions. Also evident
is that the longer the line absorber, the narrower the {}``beam''
of radiated waves and the less the energy radiated.

An expression for the full velocity potential, $\phi$, resulting
from heave motions and wave diffraction in an incident wave is obtained
by substituting the relative motion hypothesis (\ref{eq:rel_mot_hyp_far_field})
into (\ref{eq:phi_comps}) to give\begin{equation}
\phi\sim A_{0}\,\varphi_{0}+A_{r}\,\varphi_{z},\label{eq:varphi_0_z}\end{equation}
where the relative complex amplitude is defined by\begin{equation}
A_{r}\equiv A_{z}-A_{0}.\label{eq:A_r}\end{equation}
Substituting (\ref{eq:varphi_0_z}) into the integrand in the expression
for absorbed power in (\ref{eq:dE/dt_Sc_2}) and expanding gives\begin{multline}
\phi\,\frac{\partial\bar{\phi}}{\partial n}=\left|A_{0}\right|^{2}\varphi_{0}\,\frac{\partial\bar{\varphi}_{0}}{\partial n}+\bar{A}_{0}A_{r}\varphi_{z}\,\frac{\partial\bar{\varphi}_{0}}{\partial n}\\
+A_{0}\bar{A}_{r}\varphi_{0}\,\frac{\partial\bar{\varphi}_{z}}{\partial n}+\left|A_{r}\right|^{2}\varphi_{z}\,\frac{\partial\bar{\varphi}_{z}}{\partial n}.\label{eq:phi_d_phi_0}\end{multline}
The incident wave will contribute no net energy flux through the control
surface $S_{C}$, therefore $\iint_{S_{C}}\varphi_{0}\,\partial\bar{\varphi}_{0}/\partial n\, dS=0$.
With this observation the term proportional to $\left|A_{0}\right|^{2}$
in (\ref{eq:phi_d_phi_0}) can be ignored and the integral in (\ref{eq:dE/dt_Sc_2})
can be written as\begin{multline}
\iint_{S_{C}}\phi\,\frac{\partial\bar{\phi}}{\partial n}dS=\iint_{S_{C}}\left(A_{0}\bar{A}_{r}\varphi_{0}\,\frac{\partial\bar{\varphi}_{z}}{\partial n}\right.\\
\left.+\bar{A}_{0}A_{r}\varphi_{z}\,\frac{\partial\bar{\varphi}_{0}}{\partial n}+\left|A_{r}\right|^{2}\varphi_{z}\,\frac{\partial\bar{\varphi}_{z}}{\partial n}\right)dS.\label{eq:phi_d_phi}\end{multline}
The terms on the right-hand side of (\ref{eq:phi_d_phi}) are evaluated
in cylindrical polar coordinates in \ref{app:stat_phase_and_I}. The
result shows that $P$ in (\ref{eq:dE/dt_Sc_2}) can be written as\begin{multline}
P=\frac{\rho}{2}\Im\left\{ \left(A_{0}\bar{A}_{r}-\bar{A}_{0}A_{r}\right)c_{g}gk\, bl\right.\\
\left.-i\left|A_{r}\right|^{2}c_{g}gk^{3}\, b^{2}l^{2}\, I(kl)\right\} ,\label{eq:dE/dt_lineAten_0}\end{multline}
where $I(kl)$ is given by%
\footnote{The analytical expression in (\ref{eq:I_kl_analytic}) for the integral
in (\ref{eq:def_I_kl}) is given by \citeauthor{Farley82} \citeyearpar[\Eq(20)]{Farley82}.%
}\begin{alignat}{1}
I(kl) & \equiv\frac{1}{2\pi}\int_{-\pi}^{\pi}j_{0}^{2}\negmedspace\left(\frac{kl}{2}\left(1-\cos\theta\right)\right)\, d\theta,\label{eq:def_I_kl}\\
 & =\frac{4}{3}\cos(kl)\, J_{0}(kl)\nonumber \\
 & \qquad+\frac{2}{3kl}\left(2kl\sin(kl)-\cos(kl)\right)J_{1}(kl),\label{eq:I_kl_analytic}\end{alignat}
and where $J_{0}$ and $J_{1}$ are the zeroth and first-order Bessel
functions of the first kind. Since\[
A_{0}\bar{A}_{r}-\bar{A}_{0}A_{r}=2i\Im\left\{ A_{0}\bar{A}_{r}\right\} ,\]
it follows that\begin{equation}
P=\frac{1}{2}c_{g}\rho\, gk\, bl\left(2\Im\left\{ A_{0}\bar{A}_{r}\right\} -\left|A_{r}\right|^{2}k^{2}bl\, I(kl)\right).\label{eq:dEdt_lineAten_1}\end{equation}
In this equation the first term, which can be positive or negative
depending on the relative phase of $A_{0}$ and $A_{r}$, gives the
energy transferred from the fluid to the body. From the point of view
of a near-field perspective, it represents the fluid pressure on the
body times the velocity of the body; from the point of view of a far-field
perspective, it represents the interference of the incident and radiated
waves in the far-field. The second term, which is always negative,
gives the power lost due to energy radiated away from the body.

The mean power absorbed by the body, $P$ in (\ref{eq:dEdt_lineAten_1}),
is equal to the mean power extracted from the incident wave. The mean
power extracted from a wave can be written as the product of the mean
energy flux in the wave, denoted by $J$, and the power capture width,
denoted by $w$, that is\begin{equation}
P=w\, J.\label{eq:dEdt_P_wJ}\end{equation}
The mean energy flux of the plane incident wave described by (\ref{eq:varphi_0})
can be written as\begin{equation}
J=\frac{1}{2}c_{g}\rho g\left|A_{0}\right|^{2},\label{eq:J_cg_A2}\end{equation}
where $c_{g}$ is the group velocity given by\begin{equation}
c_{g}=\frac{d\omega}{dk}=\frac{\omega}{2k}\left(1+\frac{2kh}{\sinh(2kh)}\right).\label{eq:c_g}\end{equation}
Substituting (\ref{eq:dEdt_lineAten_1}) and (\ref{eq:J_cg_A2}) into
(\ref{eq:dEdt_P_wJ}) and rearranging gives the power capture width
as\begin{equation}
w=\frac{2k\, bl}{\left|A_{0}\right|^{2}}\Im\left\{ A_{0}\bar{A}_{r}\right\} -k^{3}\, b^{2}l^{2}\frac{\left|A_{r}\right|^{2}}{\left|A_{0}\right|^{2}}I(kl).\label{eq:w_p_l_1}\end{equation}

This capture width can now be optimised with respect to the phase
and amplitude of $A_{r}$. Since the second term on the right-hand
side of (\ref{eq:w_p_l_1}) is positive and independent of phase,
to optimise $w$ with respect to phase requires the maximisation of\begin{equation}
\Im\left\{ A_{0}\bar{A}_{r}\right\} =\left|A_{0}\right|\left|A_{r}\right|\sin\psi_{r}\label{eq:Im_AA_l}\end{equation}
with respect to $\psi_{r}=\arg\left(A_{0}\bar{A}_{r}\right)=\arg\left(A_{0}\right)-\arg\left(A_{r}\right)$,
which is the relative phase between $A_{0}$ and $A_{r}$. Maximisation
requires that $\sin\psi_{r}=1$, implying that $\psi_{r}=\pi/2$ so
that $A_{r}$ lags $A_{0}$ by $\pi/2$ radians and the response velocity
(which is proportional to $\partial_{n}\phi_{z}$) is in phase with
the excitation force (which is proportional to $\phi_{0}$). Substituting
$\psi_{r}=\pi/2$ into (\ref{eq:Im_AA_l}), and the result into (\ref{eq:w_p_l_1}),
gives\begin{equation}
w=2k\, bl\frac{\left|A_{0}\right|\left|A_{r}\right|}{\left|A_{0}\right|^{2}}-k^{3}\, b^{2}l^{2}\frac{\left|A_{r}\right|^{2}}{\left|A_{0}\right|^{2}}\, I(kl).\label{eq:w_p_l_2}\end{equation}
The first term in this equation, which is always positive, accounts
for the net power entering the control volume bounded by $S_{C}$
due to the interaction of the incident and radiated waves; the second
term, which is always negative, accounts for the energy leaving the
control volume due to the radiated waves.

To optimise (\ref{eq:w_p_l_2}) with respect to the relative amplitude
$\left|A_{r}\right|$ consider\[
\frac{dw}{d\left|A_{r}\right|}=2k\, bl\frac{\left|A_{0}\right|}{\left|A_{0}\right|^{2}}-2k^{3}\, b^{2}l^{2}\frac{\left|A_{r}\right|}{\left|A_{0}\right|^{2}}\, I(kl)=0,\]
the solution of which is\begin{equation}
\left|A_{r}\right|=\frac{1}{k^{2}\, bl\, I(kl)}\left|A_{0}\right|.\label{eq:Ar_l_opt_inf}\end{equation}
Substituting $\left|A_{r}\right|$ from (\ref{eq:Ar_l_opt_inf}) into
the first and second terms on the right-hand side of (\ref{eq:w_p_l_2})
gives the optimal amplitude condition that the radiated power represented
by the second term is exactly half of the absorbed power represented
by the first term. This, together with the optimal phase condition
that $A_{r}$ lags $A_{0}$ by $\pi/2$ radians corresponds to the
familiar condition of {}``impedance matching'' to maximise the energy
transfer between a source and a load.

The value of $\left|A_{r}\right|$ in (\ref{eq:Ar_l_opt_inf}) is
not limited by the volume of the device, and since $k$ is typically
relatively small the value of $\left|A_{r}\right|$ can be very large
in comparison with $\left|A_{0}\right|$. Because of this, when $\left|A_{r}\right|$
from (\ref{eq:Ar_l_opt_inf}) is substituted back into (\ref{eq:w_p_l_2})
it gives the optimum capture width for a device of unconstrained motion,
or equivalently, unlimited volume, as\begin{equation}
w=\frac{1}{k\, I(kl)}\quad\mbox{if }\left|A_{r}\right|\mbox{ is unlimited}.\label{eq:w_l_inf}\end{equation}
This is identical to the result reported by \citet{Farley82} (see
\Eq(18) and the appendix of that work).

In the limit that the length of the line absorber tends to zero, by
(\ref{eq:I_limit}) in \ref{app:stat_phase_and_I}, $\lim_{kl\to0}I(kl)\to1$
and therefore $w=1/k$, which agrees with the result obtained for
the capture width of an unconstrained heaving point absorbed by, for
example, \citet{Evans1976} and \citet{Newman1976}.

At this point a physical interpretation can be given as to why higher
capture widths are achievable from line absorbers than from point
absorbers: for a line absorber the energy being carried away by radiated
waves can be made arbitrarily small while maintaining the amplitude,
$\left|A_{r}\right|$, of the radiated wave in the $\theta=0$ direction
that is necessary to remove energy from the incident wave by destructive
interference. This interpretation is clear from consideration of (\ref{eq:w_p_l_2})
in the light of (\ref{eq:def_I_kl}) and \F\ref{fig:j_0}. Obviously,
to increase the capture width the energy loss given by the second
term on the right-hand side of (\ref{eq:w_p_l_2}) should be minimised
while maintaining a high value for the energy absorption given by
the first term on the right-hand side. For a point absorber $I(kl)=1$,
so $\left|A_{r}\right|$ is the only parameter available that is independent
of the incident wave and whose value can be adjusted to maximise $w$.
The magnitude of $\left|A_{r}\right|$ needs to be increased to increase
the energy absorption from the first term in (\ref{eq:w_p_l_2}),
which is linear in $\left|A_{r}\right|$; but if $\left|A_{r}\right|$
is made too large the energy loss from the second term in (\ref{eq:w_p_l_2}),
which is quadratic in $\left|A_{r}\right|$, dominates. For a line
absorber, however, the added degree-of-freedom of length can be used
to reduce the spread of the radiated waves, as shown in \F\ref{fig:j_0}.
This reduces the magnitude of $I(kl)$ while keeping a large magnitude
for $\left|A_{r}\right|$, so achieving a high energy absorption from
the first term on the right-hand side of (\ref{eq:w_p_l_2}) and a
low energy loss from the second.

\subsection{Capture width for limited volume}

In the preceding section no constraints were placed on the maximum
value of the relative amplitude, $\left|A_{r}\right|$, and the maximum
capture width for a line absorber of unlimited volume was derived
as (\ref{eq:w_l_inf}). In this section a motion constraint is placed
on $\left|A_{r}\right|$ in (\ref{eq:w_p_l_2}). In the linear theory
presented here such a constraint on $\left|A_{r}\right|$ may be equated
to a limit on the maximum swept volume of the device. This constraint
may be a consequence of the limited volume of a device, or it may
be a consequence of other engineering or control constraints that
limit the range of heave motion. Assume for simplicity that the depth,
$D$, and the beam, $b$, of each vertically-wall-sided element are
independent of its position along the length, $l$, of the line absorber.
Here, $D$ is the vertical difference in position between the minimum
and maximum submergences of each element of the line absorber. If
the element is symmetric about the plane $z=0$ the draft and the
freeboard are equal and their sum is equal to $D$. In this case half
the maximum swept volume of the line absorber is given by\[
V=bl\,\frac{D}{2},\]
and the maximum value of $\left|A_{r}\right|$ is\begin{equation}
\max\left(\left|A_{r}\right|\right)=\frac{D}{2}=\frac{V}{bl}.\label{eq:Ar_l_max}\end{equation}
Substituting $\left|A_{r}\right|$ from (\ref{eq:Ar_l_opt_inf}) into
(\ref{eq:Ar_l_max}) gives the volume limiting condition in terms
of $\left|A_{0}\right|$ as\begin{equation}
\max\left(\left|A_{0}\right|\right)=k^{2}V\, I(kl).\label{eq:A0_max-1}\end{equation}
Thus, for a limited volume device the maximum $\left|A_{r}\right|$
is reached when $\left|A_{0}\right|>k^{2}V\, I(kl)$ and the maximum
capture width is given by substituting the maximum of $\left|A_{r}\right|$
from (\ref{eq:Ar_l_max}) into (\ref{eq:w_p_l_2}). This gives\begin{multline*}
w=\frac{kV}{\left|A_{0}\right|}\left(2-\frac{k^{2}V}{\left|A_{0}\right|}\, I(kl)\right)\\
\quad\mbox{if }\left|A_{0}\right|>k^{2}V\, I(kl).\end{multline*}
When $\left|A_{0}\right|\le k^{2}V\, I(kl)$ the optimum capture width
is still given by (\ref{eq:w_l_inf}). Thus, the maximum capture width
of a volume-limited heaving line attenuator can be written as\begin{equation}
w=\left\{ \begin{array}{cc}
\negthickspace\negthickspace{\displaystyle \frac{1}{k\, I(kl)}}, & {\displaystyle \negthickspace\left|A_{0}\right|\le k^{2}V\, I(kl)},\\
\\\negthickspace\negthickspace{\displaystyle \frac{kV}{\left|A_{0}\right|}\negthickspace\left(2-\frac{k^{2}V}{\left|A_{0}\right|}\, I(kl)\right)}, & {\displaystyle \negthickspace\left|A_{0}\right|>k^{2}V\, I(kl),}\end{array}\right.\label{eq:w_l_V}\end{equation}
where an analytic expression for $I(kl)$ is given by (\ref{eq:I_kl_analytic}).

\section{Dimensionless capture width and absorbed power}

\begin{figure*}
\noindent \begin{centering}
\includegraphics[width=0.5\textwidth]{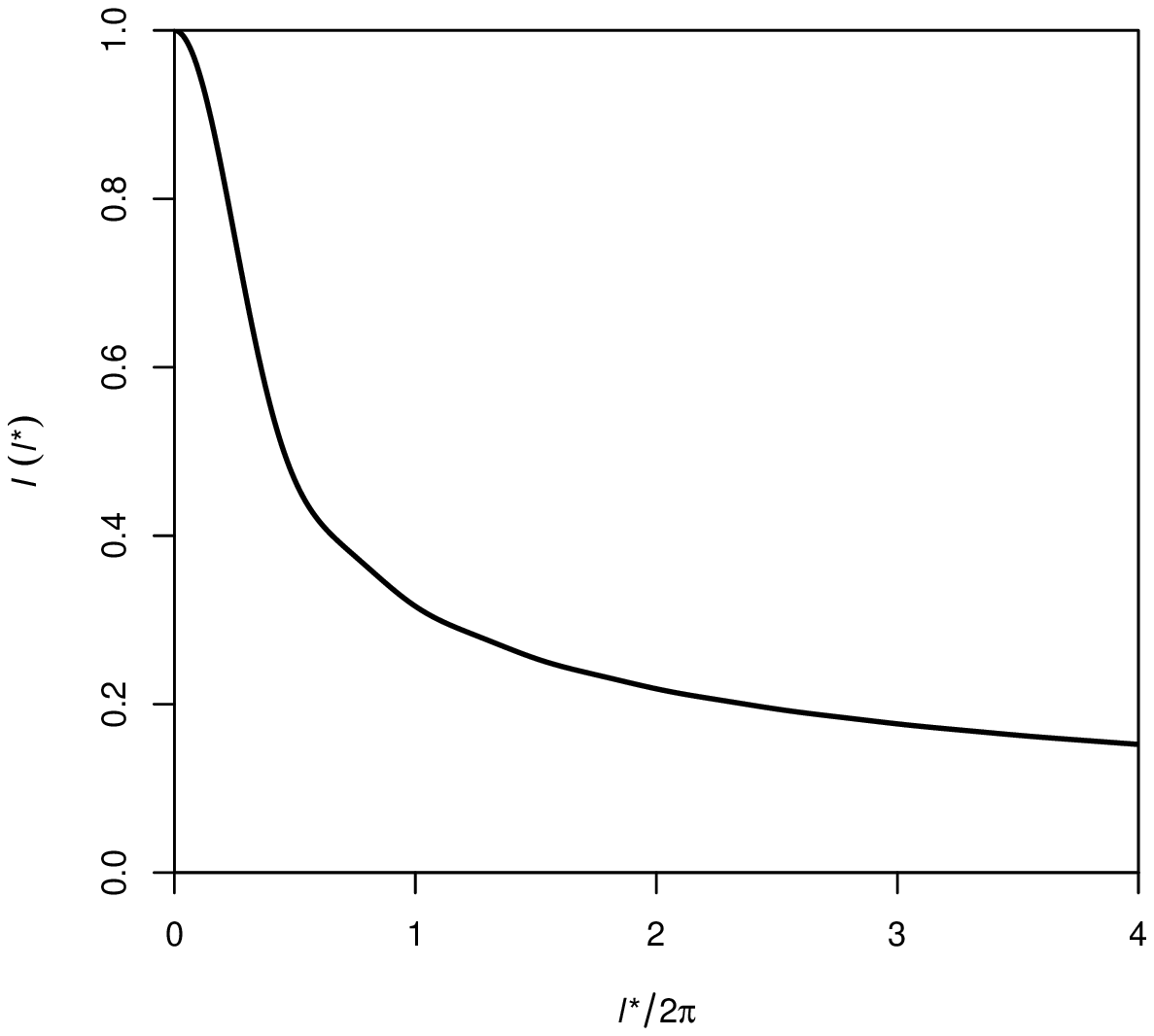}\includegraphics[width=0.5\textwidth]{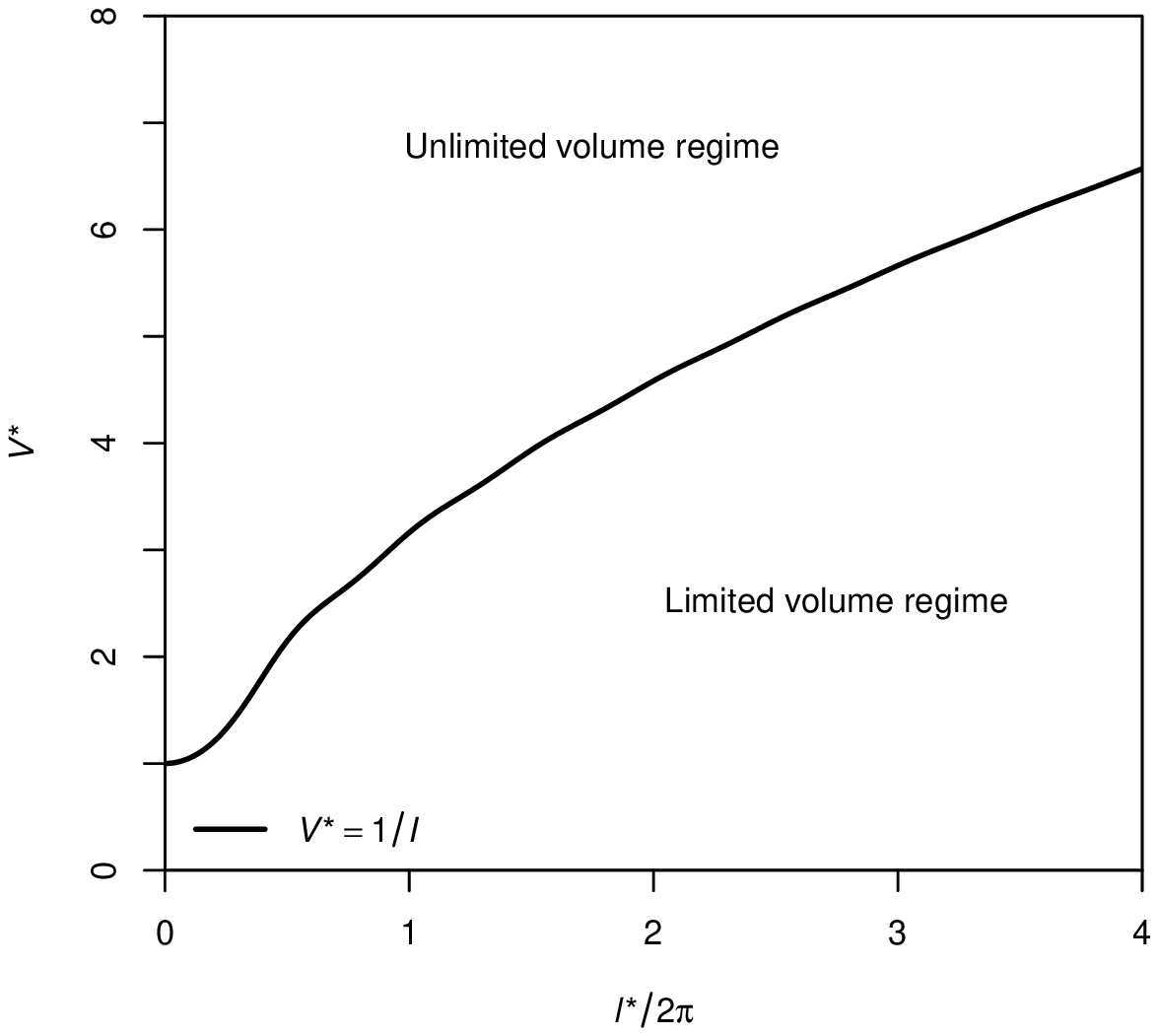}
\par\end{centering}

\caption{\label{fig:fig3a.eps}The dimensionless functions $I\left(kl\right)=I\left(l^{*}\right)$
and $V^{*}=1/I\left(l^{*}\right)$. }
\ifthenelse{\boolean{includeNotes}}{

\textcolor{green}{Figures produced by \textasciitilde{}/opd/documents/line\_absorbers/R-analysis/power\_plots\_AOR.R.}

}

\end{figure*}

It is instructive to formulate the results of \S\ref{sec:attenuator}
in dimensionless units. The dimensionless independent variables of
device length and device half-swept volume, and the dimensionless
dependent variable of capture width, are defined by \begin{eqnarray}
l^{*} & \equiv & kl,\label{eq:l^*}\\
V^{*} & \equiv & \frac{k^{2}V}{\left|A_{0}\right|},\label{eq:V^*}\\
w^{*} & \equiv & kw.\label{eq:w^*}\end{eqnarray}
Since capture width and absorbed power are related by (\ref{eq:dEdt_P_wJ}),
the definition of $w^{*}$ in (\ref{eq:w^*}) suggests a non-dimensional
power defined by\begin{equation}
P^{*}\equiv\frac{k\, P}{J},\label{eq:P^*}\end{equation}
where $J$ is given by (\ref{eq:J_cg_A2}). From (\ref{eq:dEdt_P_wJ}),
(\ref{eq:w^*}) and (\ref{eq:P^*}) is follows that the dimensionless
variables $w^{*}$ and $P^{*}$ are equivalent. Substituting (\ref{eq:l^*})--(\ref{eq:w^*})
into (\ref{eq:w_l_V}) gives the dimensionless capture width and power
as\begin{equation}
w^{*}\equiv P^{*}=\left\{ \begin{array}{cc}
\negthickspace{\displaystyle \frac{1}{I\left(l^{*}\right)}}, & {\displaystyle V^{*}\, I\left(l^{*}\right)\ge1,}\\
\\\negthickspace V^{*}{\displaystyle \left(2-V^{*}\, I\left(l^{*}\right)\right)}, & {\displaystyle V^{*}\, I\left(l^{*}\right)<1,}\end{array}\right.\label{eq:w_l_V^*}\end{equation}
where $I\left(l^{*}\right)=I(kl)$ is the integral given by (\ref{eq:I_kl_analytic}).
The forms of the functions $I\left(l^{*}\right)$ and $1/I\left(l^{*}\right)$
are shown in \F\ref{fig:fig3a.eps}. In these plots $l^{*}/2\pi=l/\lambda$
is used as the dimensionless unit on the abscissa as these units correspond
to multiples of the wavelength. Shown in the right-hand side plot
in \F\ref{fig:fig3a.eps} is the line $V^{*}=1/I\left(l^{*}\right)$
which marks the boundary between the limited and unlimited volume
regimes.

The point absorber limit is obtained by substituting $\lim_{l^{*}\to0}I\left(l^{*}\right)\to1$
into (\ref{eq:w_l_V^*}) to give\begin{equation}
\lim_{l^{*}\to0}w^{*}\to\left\{ \begin{array}{cc}
{\displaystyle 1}, & {\displaystyle V^{*}\ge1,}\\
\\2V^{*}{\displaystyle -V^{*2}}, & {\displaystyle V^{*}<1.}\end{array}\right.\label{eq:w_p_V^*}\end{equation}
The term in (\ref{eq:w_p_V^*}) which applies when ${\displaystyle V^{*}\ge1}$,
that is $w^{*}=1$, is the well known result for the theoretical maximum
capture width of a heaving point absorber of unlimited volume. When
${\displaystyle V^{*}\ll1}$ the term, which is first-order in volume,
dominates and the capture width can be written as $w^{*}=2V^{*}$,
or, equivalently, $P^{*}/V^{*}=2$. This corresponds to Budal's upper
bound%
\footnote{Budal's result is generally quoted as the upper bound on the absorbed
power per unit volume given by\[
\frac{P}{V_{T}}=\frac{1}{4}\rho g\omega\left|A_{0}\right|,\]
where $V_{T}=2V$ is the total swept volume of the body.%
} on the maximum capture width, or power per unit volume, of a finite-volume
heaving point absorber \citep[see][]{Falnes1993}. The term that is
second-order in volume is a modification arising from the inclusion
of energy loss due to radiated waves. 

The infinite length line absorber limit is obtained by substituting
$\lim_{l^{*}\to\infty}I\left(l^{*}\right)\to0$ into (\ref{eq:w_l_V^*})
to give\begin{equation}
\lim_{l^{*}\to\infty}w^{*}\to2V^{*}\quad\mbox{for all }V^{*}.\label{eq:w_l_V^*_l-infty}\end{equation}
This can also be written as $P^{*}/V^{*}=2$ for all $V^{*}$. Comparing
(\ref{eq:w_p_V^*}) and (\ref{eq:w_l_V^*_l-infty}) shows that for
a heaving point absorber the maximum capture width is limited to $w^{*}=1$,
no matter how great the volume; for a line absorber in the limit infinite
length, however, the maximum capture width scales linearly with volume,
that is $w^{*}=2V^{*}$. 

\begin{figure*}
\noindent \begin{centering}
\includegraphics[width=0.5\textwidth]{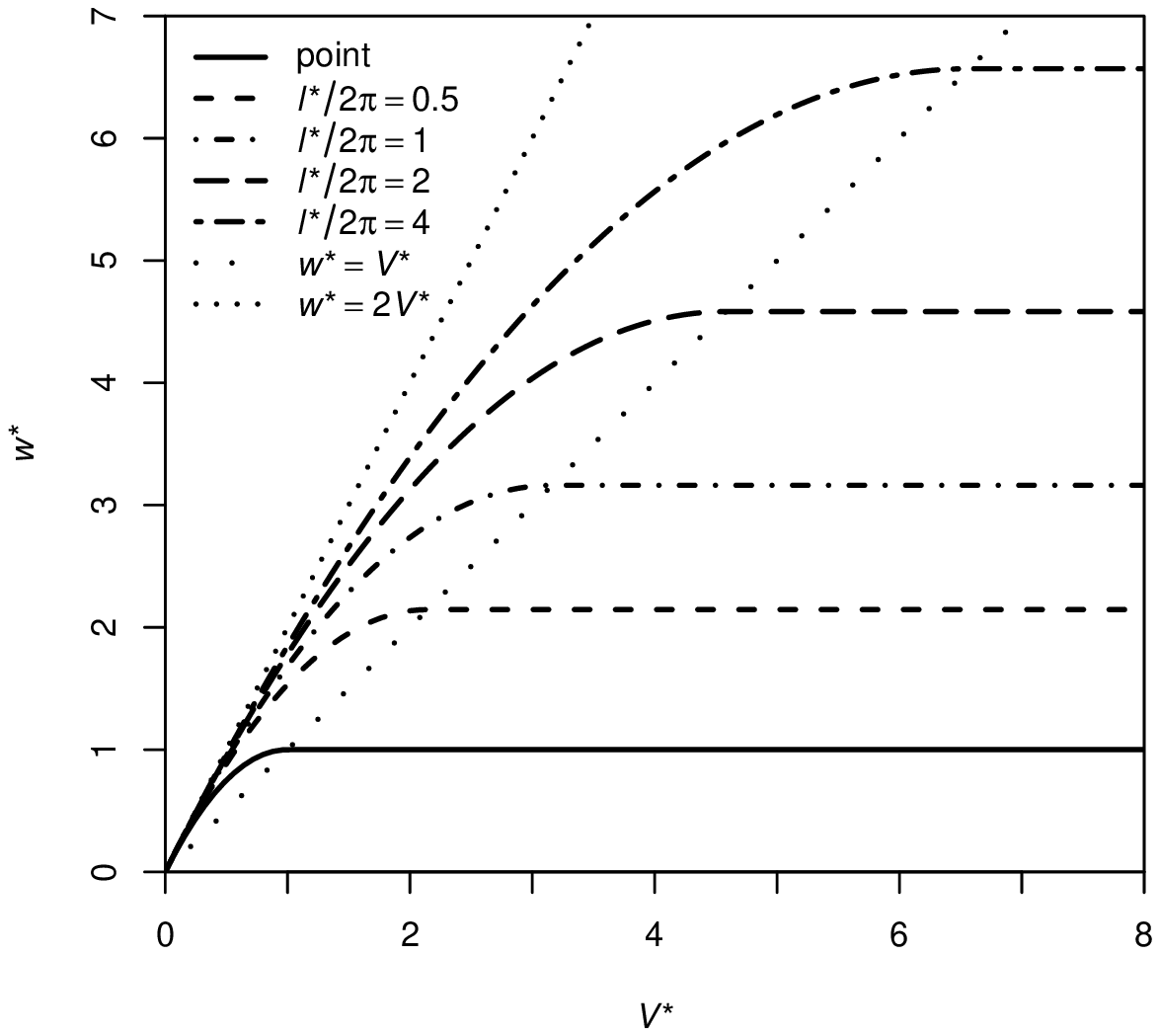}\includegraphics[width=0.5\textwidth]{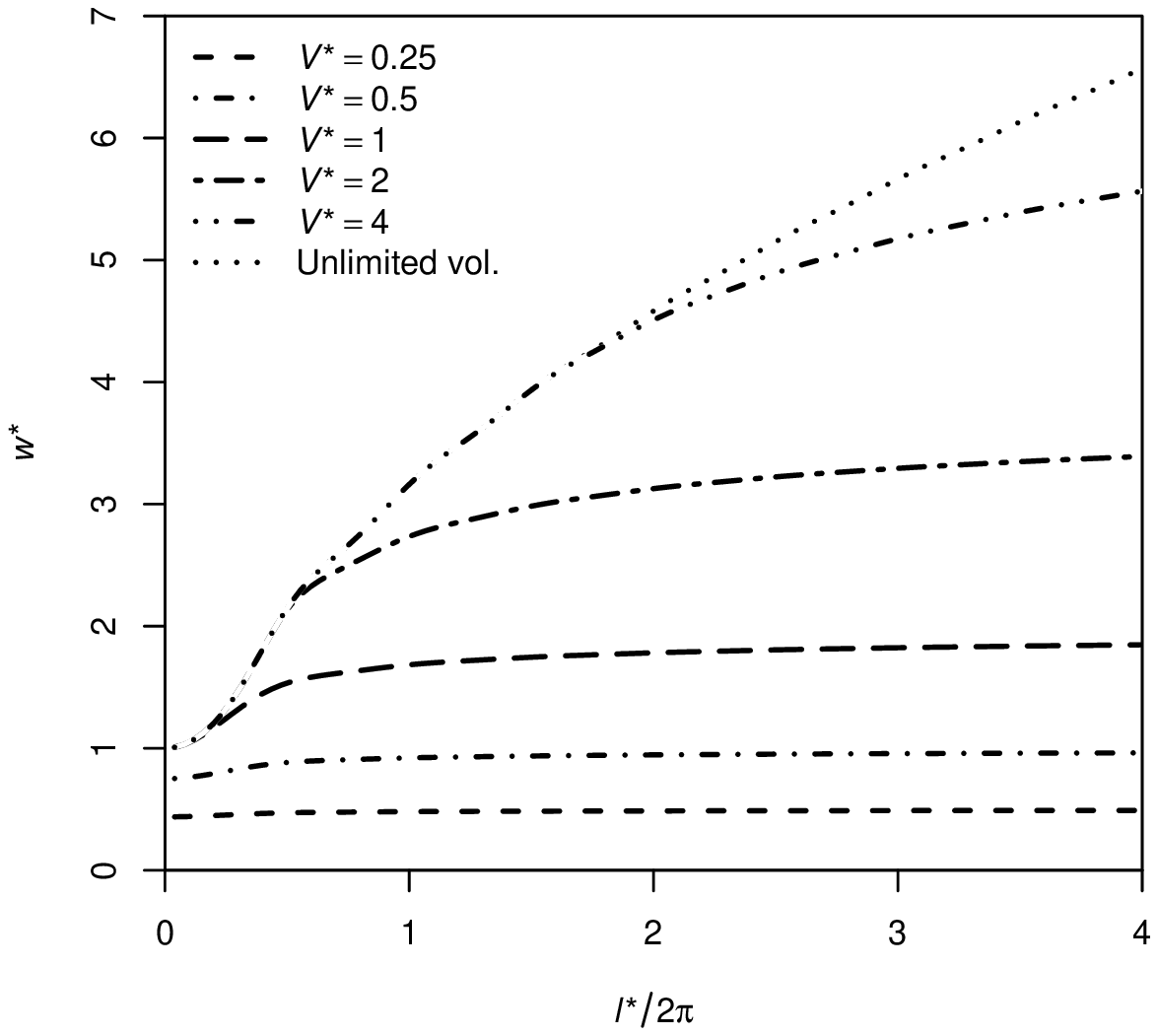}
\par\end{centering}

\caption{\label{fig:w^*}Dimensionless capture width, $w^{*}$, from (\ref{eq:w_l_V^*})
as a function of dimensionless half-swept volume, $V^{*}$, and dimensionless
device length, $l^{*}$.}
\ifthenelse{\boolean{includeNotes}}{

\textcolor{green}{Figures produced by \textasciitilde{}/opd/documents/line\_absorbers/R-analysis/power\_plots\_AOR.R.}

}

\end{figure*}

Plots of dimensionless capture width (or absorbed power) from (\ref{eq:w_l_V^*})
are shown in \F\ref{fig:w^*}. The two plots give different illustrations
of the effect on $w^{*}$ of changing the volume and length of a device.
The solid line in the left-hand plot in \F\ref{fig:w^*} shows that
increasing volume above $V^{*}=1$ gives no increase in $w^{*}$ for
a point absorber; the dashed lines show that, for line absorbers,
increasing $V^{*}$ above $1$ can increase $w^{*}$, and that the
increase is greater for a longer device. In fact, the dotted line
marks where $w^{*}=2V^{*}$ which shows that the maximum dimensionless
capture widths for an infinite length line absorber is equal to its
dimensionless swept volume. The fine dotted line marks where $w^{*}=V^{*}$
which is the line of transition between the unlimited volume regime
below the line and the limited volume regime above it. It is clear
from this plot that the minimum half-swept volume required for a point
absorber to achieve its maximum capture width is $V^{*}=1$, and for
a line absorber it is $V^{*}=w^{*}$. Note that $V^{*}=1$ is equivalent
to $V/\lambda^{2}\left|A_{0}\right|=1/4\pi^{2}\approx0.025$, which
aids in interpreting the dimensionless volume as it states that the
minimum half-swept volume for a point absorber to achieve its maximum
capture width is about $2.5\%$ of the natural measure of volume given
by the square of the wavelength times the amplitude of the wave, that
is, $\lambda^{2}\left|A_{0}\right|$.

The right-hand plot in \F\ref{fig:w^*} shows the gains in $w^{*}$
that can be achieved by {}``stretching out'' different fixed-volume
point absorbers to line absorbers of different lengths. The points
where the lines terminate at $l^{*}=0$ represent the dimensionless
capture widths of point absorbers of the appropriate volumes. There
is little difference in $w^{*}$ between point and line absorbers
when $V^{*}<0.5$, that is, well within the volume-limited regime
of the point absorber. For the unlimited volume regime of $V^{*}\ge1$
the capture width of the point absorber is at its theoretical maximum
of $w^{*}=1$ and the lines for $V^{*}=1$, $2$ and $4$ all converge
to $w^{*}=1$ as $l^{*}\to0$. For finite length line absorbers, however,
significant increases in capture width are achievable above $w^{*}=1$,
as shown in the figure and in the examples given in \T\ref{tab:V_l_w}.

\begin{table}
\noindent \begin{centering}
\begin{tabular}{ccc}
\hline 
half-swept  & length,  & capture-\tabularnewline
volume, $V^{*}$  & $l^{*}/2\pi=l/\lambda$  & width, $w^{*}$\tabularnewline
\hline 
1 & 0 & 1\tabularnewline
1 & 1 & 1.684\tabularnewline
1 & 2 & 1.782\tabularnewline
2 & 0 & 1\tabularnewline
2 & 1 & 2.735\tabularnewline
2 & 2 & 3.127\tabularnewline
3 & 0 & 1\tabularnewline
3 & 1 & 3.154\tabularnewline
3 & 2 & 4.036\tabularnewline
unlimited & 0 & 1\tabularnewline
unlimited & 1 & 3.162\tabularnewline
unlimited & 2 & 4.583\tabularnewline
\hline
\end{tabular}\caption{\label{tab:V_l_w}Examples of maximum dimensionless capture widths
for different dimensionless volumes and lengths.}

\par\end{centering}

\noindent \ifthenelse{\boolean{includeNotes}}{

\textcolor{green}{Table numbers produced by \textasciitilde{}/opd/documents/line\_absorbers/R-analysis/power\_plots\_AOR.R.}

}

\end{table}

\begin{figure*}
\noindent \begin{centering}
\includegraphics[width=0.5\textwidth]{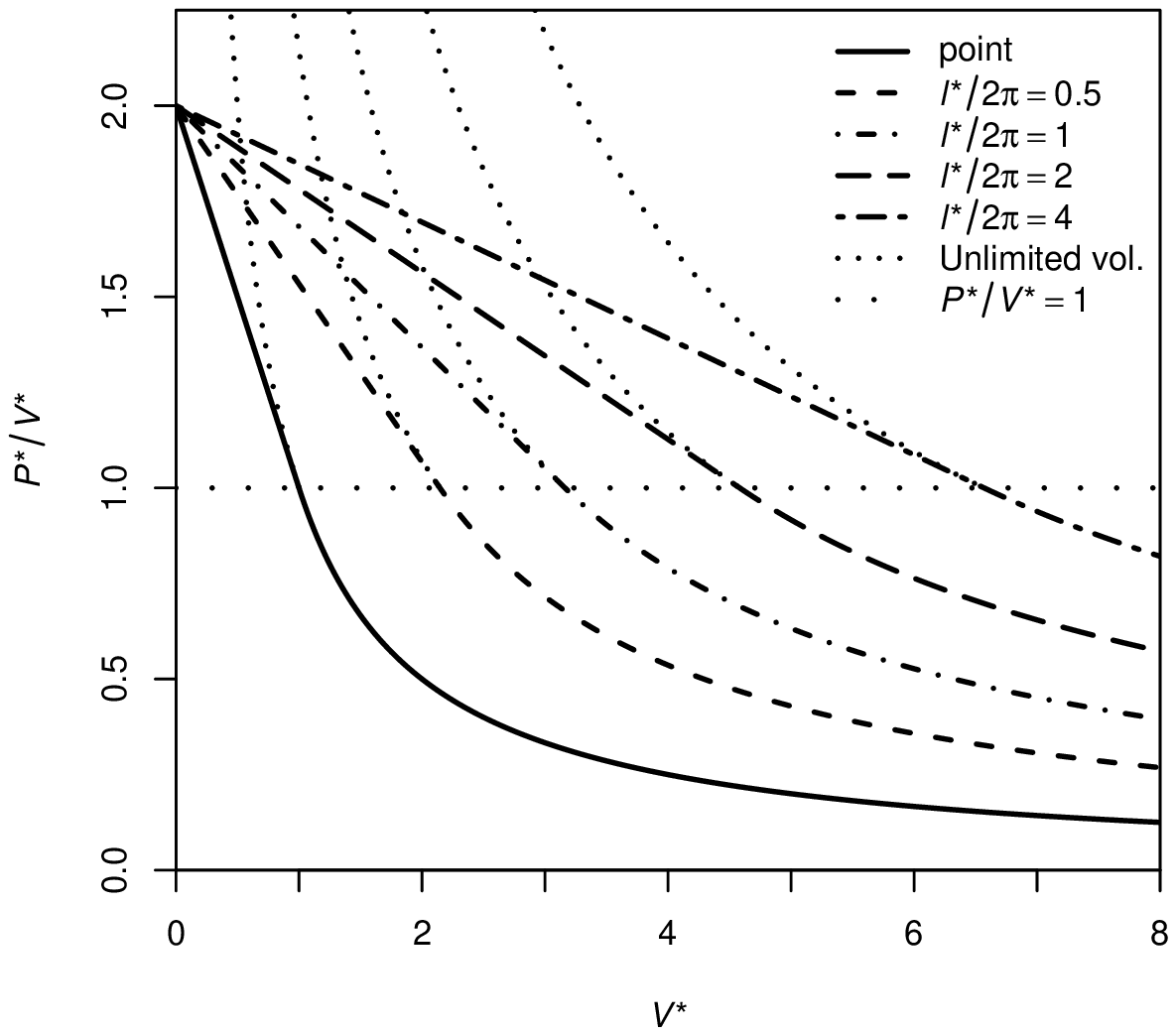}\includegraphics[width=0.5\textwidth]{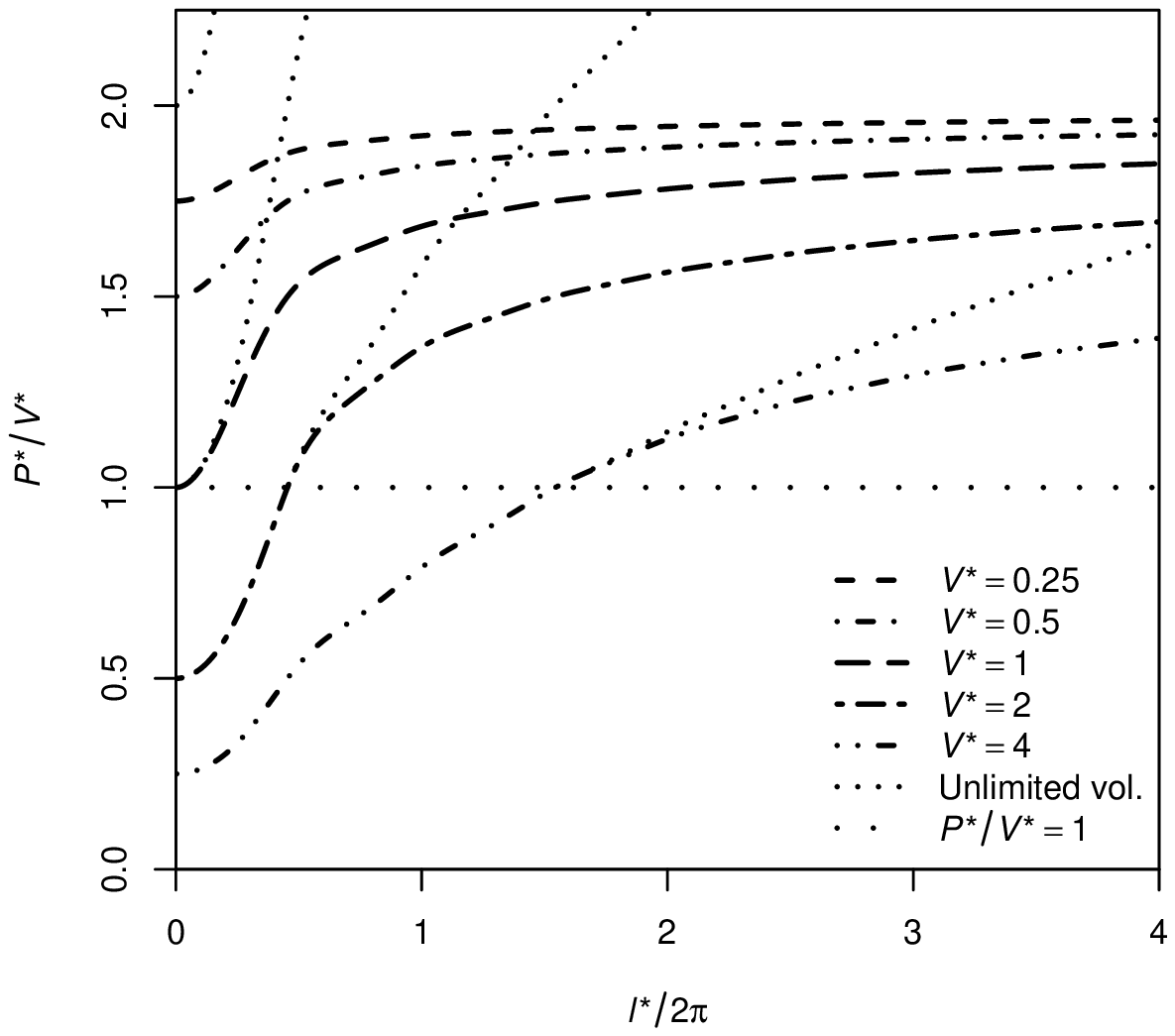}
\par\end{centering}

\caption{\label{fig:P^*}Dimensionless power per unit dimensionless volume,
$P^{*}/V^{*}$, from (\ref{eq:w_l_V^*}) as a function of dimensionless
half-swept volume, $V^{*}$, and dimensionless device length, $l^{*}$.}
\ifthenelse{\boolean{includeNotes}}{

\textcolor{green}{Figures produced by \textasciitilde{}/opd/documents/line\_absorbers/R-analysis/power\_plots\_AOR.R.}

}

\end{figure*}

Plots of dimensionless absorbed powers (or capture widths) per unit
dimensionless volume are shown in \F\ref{fig:P^*}. In the left-hand
plot in \F\ref{fig:P^*} the fine dotted line marks the line of $P^{*}/V^{*}=1$,
which is the line of transition between the limited volume regime
above the line and unlimited volume regime below it. In this plot
the limited-volume dimensionless powers converge to $P^{*}/V^{*}=2$
at $V^{*}=0$ for all device lengths $l^{*}$. This maximum value
of $P^{*}/V^{*}=2$ corresponds to Budal's upper bound expressed in
the dimensional units of (\ref{eq:l^*})--(\ref{eq:P^*}). Budal's
upper bound applies to all wave energy absorbers, however, point absorbers
can only reach Budal's upper bound in the zero-volume limit, whereas
line absorber in the infinite-length limit can reach Budal's upper
bound for any volume. This is evident from the infinite-length limit
of (\ref{eq:w_l_V^*}) giving $P^{*}/V^{*}=2$ which is independent
of volume. From the left-hand plot in \F\ref{fig:P^*} it is clear
that for small volume devices there is no significant advantage to
using an attenuating line absorber compared to a point absorber. For
larger volume devices, however, the same plot shows how the power
per unit volume decreases less rapidly for a line absorber than for
an equivalent volume point absorber, and the rate of decrease of power
per unit volume is less for longer line absorbers (being zero in the
limit of an infinitely long line absorber). In fact, by dividing volume
of a single point absorber with $V^{*}=1$ into ever smaller volume
point absorbers the maximum power per unit volume that can be achieved
goes from $P^{*}/V^{*}=1$ to $P^{*}/V^{*}=2$ and, therefore, the
maximum absorbed power is doubled. This severe drop-off in $P^{*}/V^{*}$
with increasing $V^{*}$ for point absorbers is an illustration of
the {}``small is beautiful'' argument made by \citet{Falnes1993}.

The right-hand plot in \F\ref{fig:P^*} gives an alternative illustration
of how, if efficiency is measured in terms of absorbed power per unit
volume, for all device lengths, $l^{*}$, devices of smaller volume
always have greater efficiencies. Note that in this plot the dotted
line starting at $l^{*}=0$ and $P^{*}/V^{*}=2$ represents the power
per unit volume in the unlimited volume regime for the case $V^{*}=0.5$.
The corresponding line for $V^{*}=0.25$ is off the scale.

\section{Application to realistically sized WECs}

\begin{table}
\noindent \centering{}\begin{tabular}{ccc}
\hline 
half-swept  &  & \tabularnewline
volume, & length,  & \tabularnewline
$V$ (m$^{3}$)  & $l$ (m)  & WEC of similar size\tabularnewline
\hline 
240 & 0 & OPT PB150 \citep{pb150}\tabularnewline
940 & 0 & WaveBob \citep{wavebob}\tabularnewline
350 & 120 & Pelamis FSP \citep{pelamisFSP}\tabularnewline
790 & 180 & Pelamis P2 \citep{pelamisP2}\tabularnewline
1700 & 210 & possible future Pelamis\tabularnewline
\hline
\end{tabular}\caption{\label{tab:devices}Selection of WEC sizes chosen to represent realistic
production sizes based on estimates of the sizes of devices already
built or recently proposed.}

\end{table}

The advantage of formulating the results in dimensionless units is
that the number of independent variables is reduced from four ($k$,
$l$, $V$, and $\left|A_{0}\right|$) to just two ($l^{*}$ and $V^{*}$).
In dimensionless units, however, it is not immediately obvious where
a particular incident wave and WEC combination will occur on the graphs.
In this section we consider the five realistically sized WECs described
in \T\ref{tab:devices}. It is stressed that the results in the following
figures are not representative of the actual capture widths or absorbed
powers of these devices, but only that devices of these approximate
volumes and lengths have already been built or may be built in the
near future.

\begin{figure*}
\noindent \begin{centering}
\includegraphics[width=0.5\textwidth]{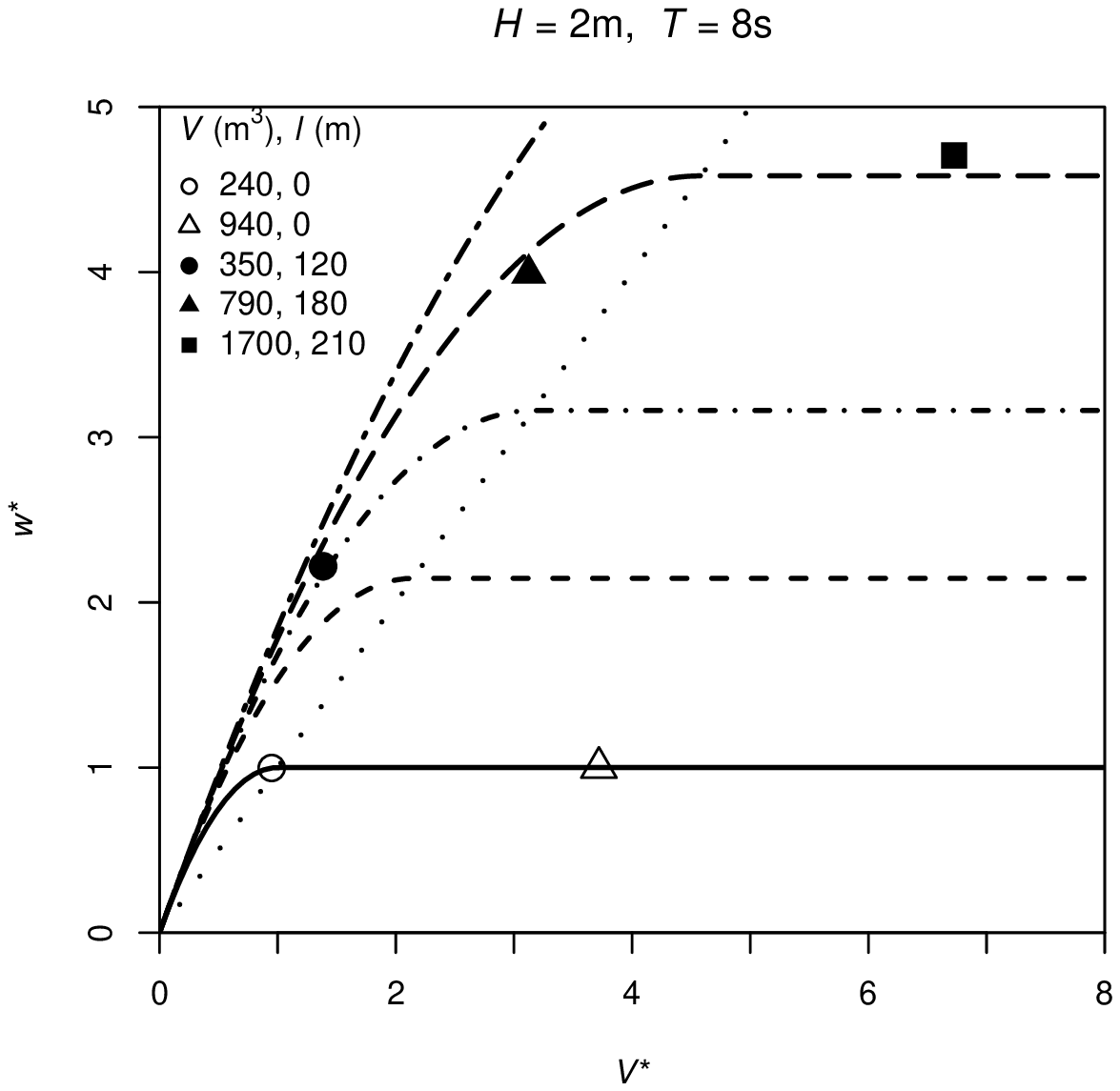}\includegraphics[width=0.5\textwidth]{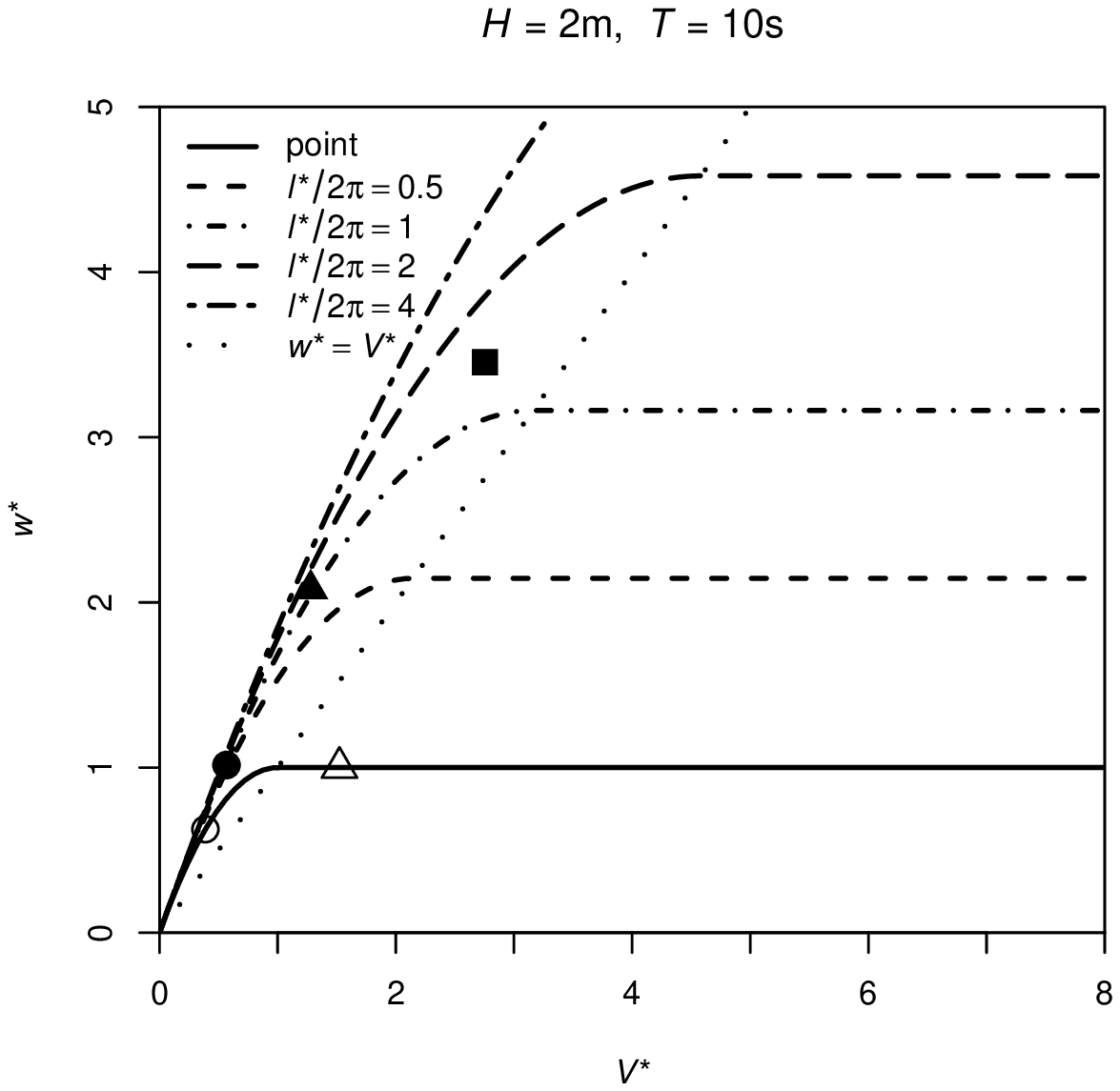}
\par\end{centering}

\noindent \begin{centering}
\includegraphics[width=0.5\textwidth]{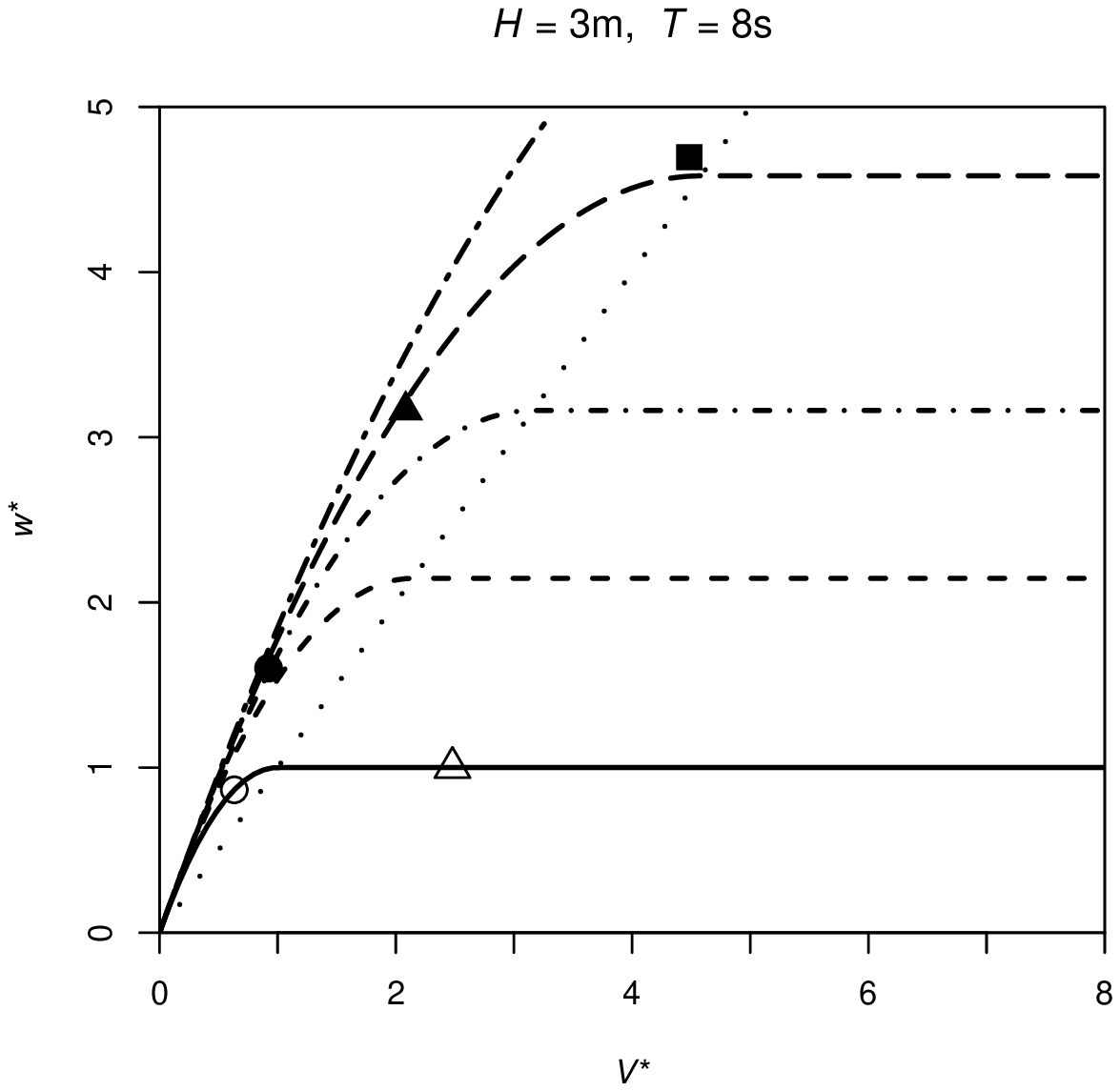}\includegraphics[width=0.5\textwidth]{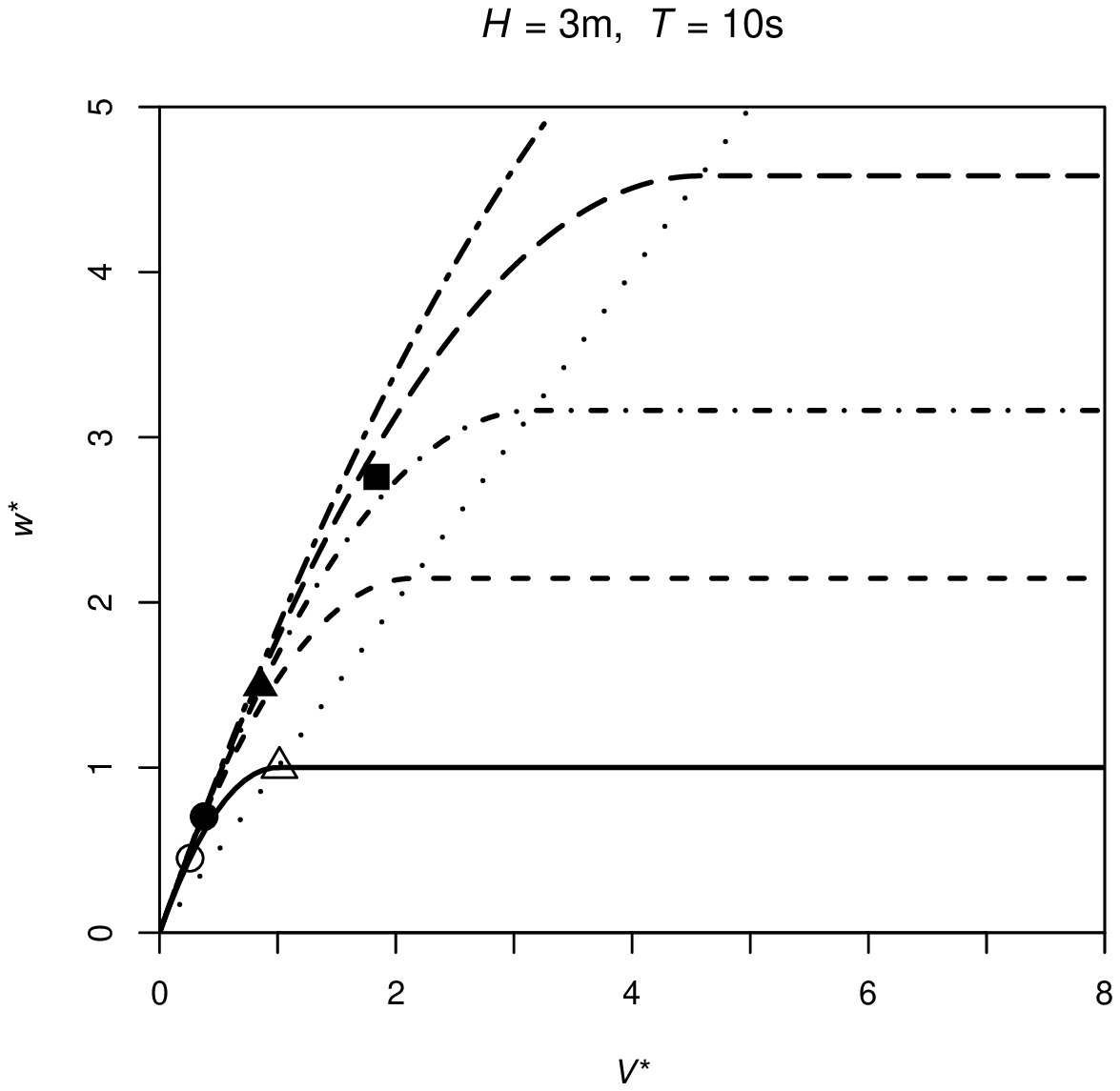}
\par\end{centering}

\caption{\label{fig:w^*_V^*_devs}Dimensionless capture width, $w^{*}$, from
(\ref{eq:w_l_V^*}) as a function of dimensionless half-swept volume,
$V^{*}$, for WECs in \T\ref{tab:devices} operating in four different
incident wave defined by their $(H,T)$.}
\ifthenelse{\boolean{includeNotes}}{

\textcolor{green}{Figures produced by \textasciitilde{}/opd/documents/line\_absorbers/R-analysis/power\_plots\_AOR.R.}

}

\end{figure*}

\begin{figure*}
\noindent \begin{centering}
\includegraphics[width=0.5\textwidth]{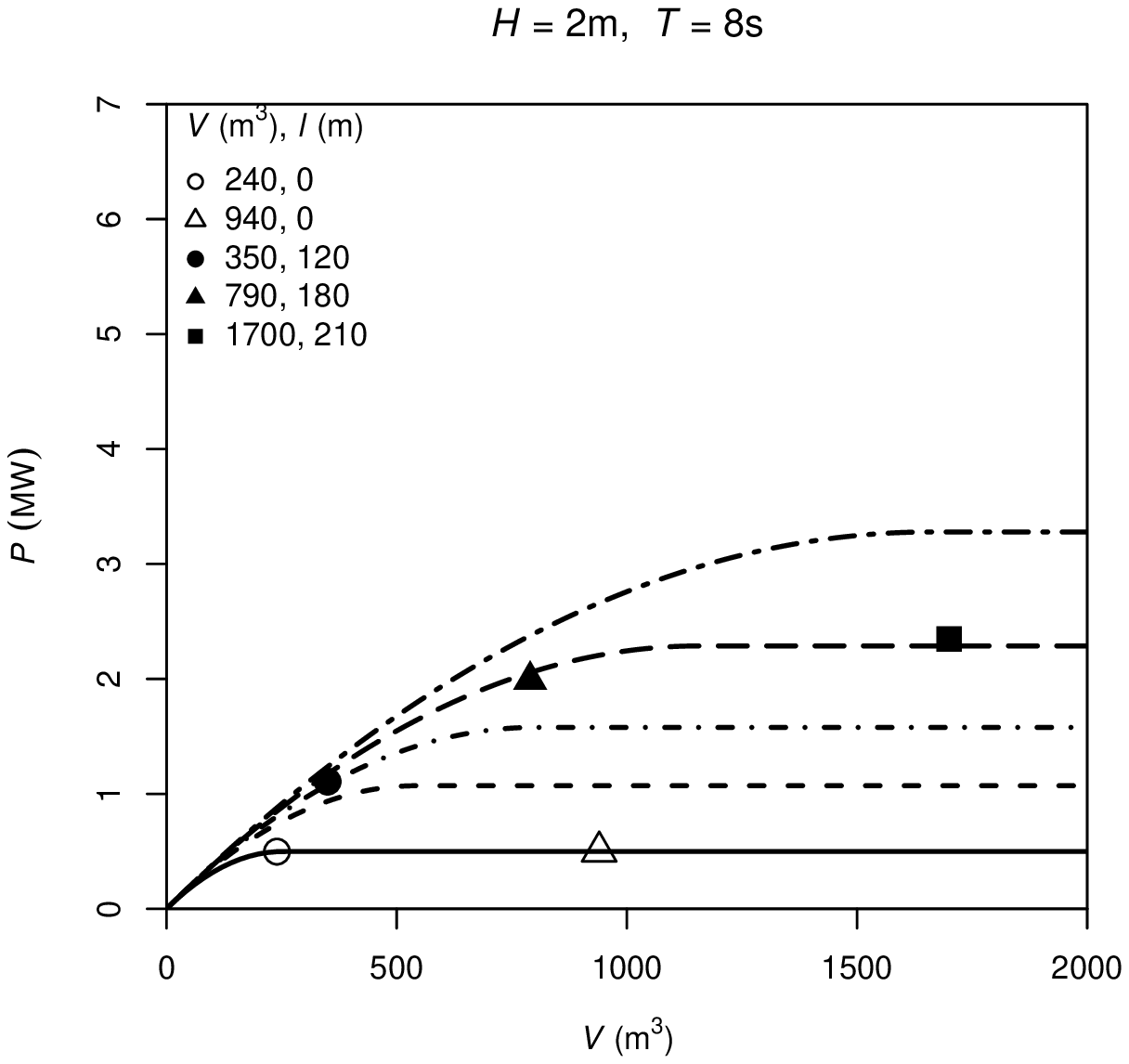}\includegraphics[width=0.5\textwidth]{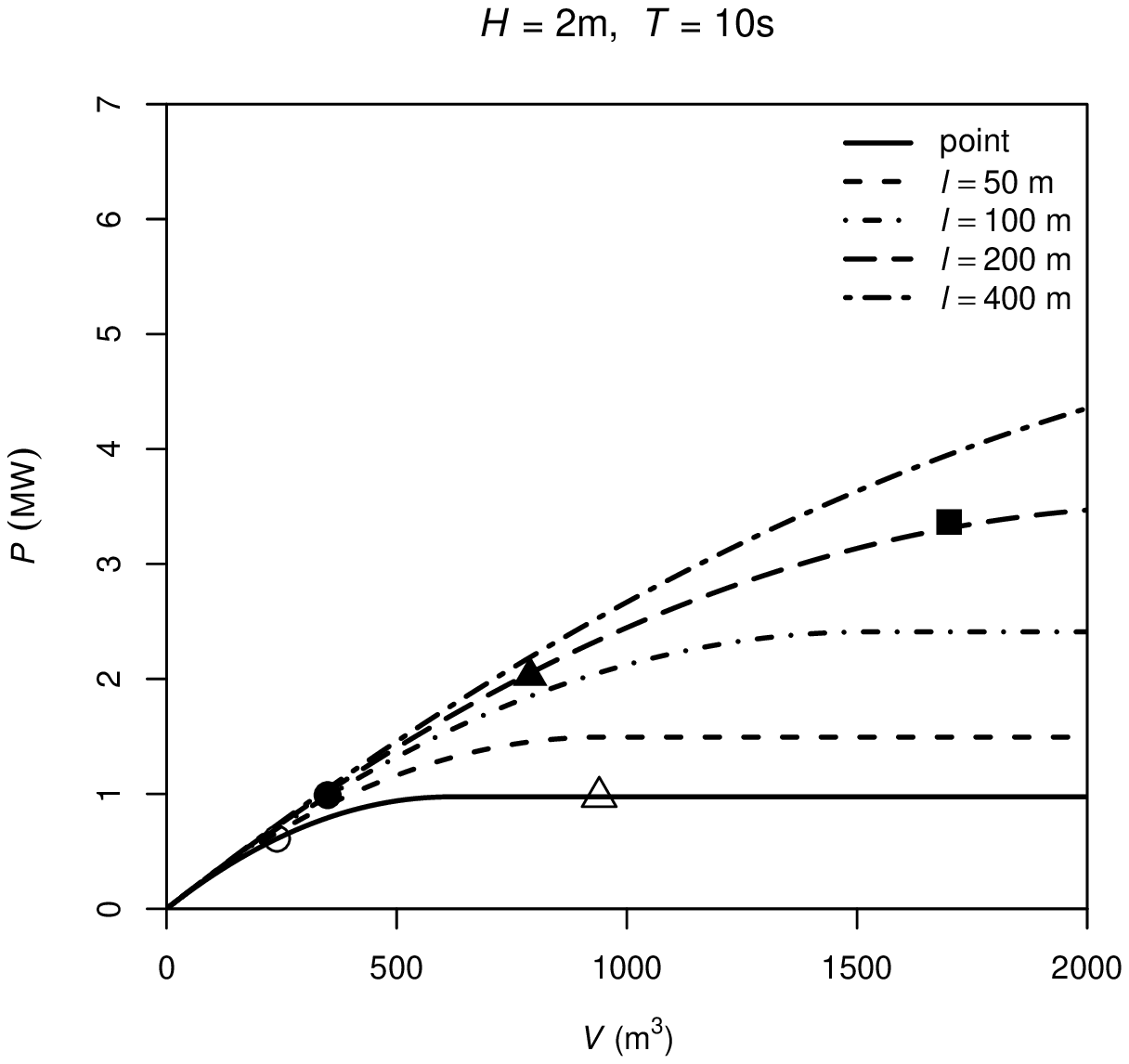}
\par\end{centering}

\noindent \begin{centering}
\includegraphics[width=0.5\textwidth]{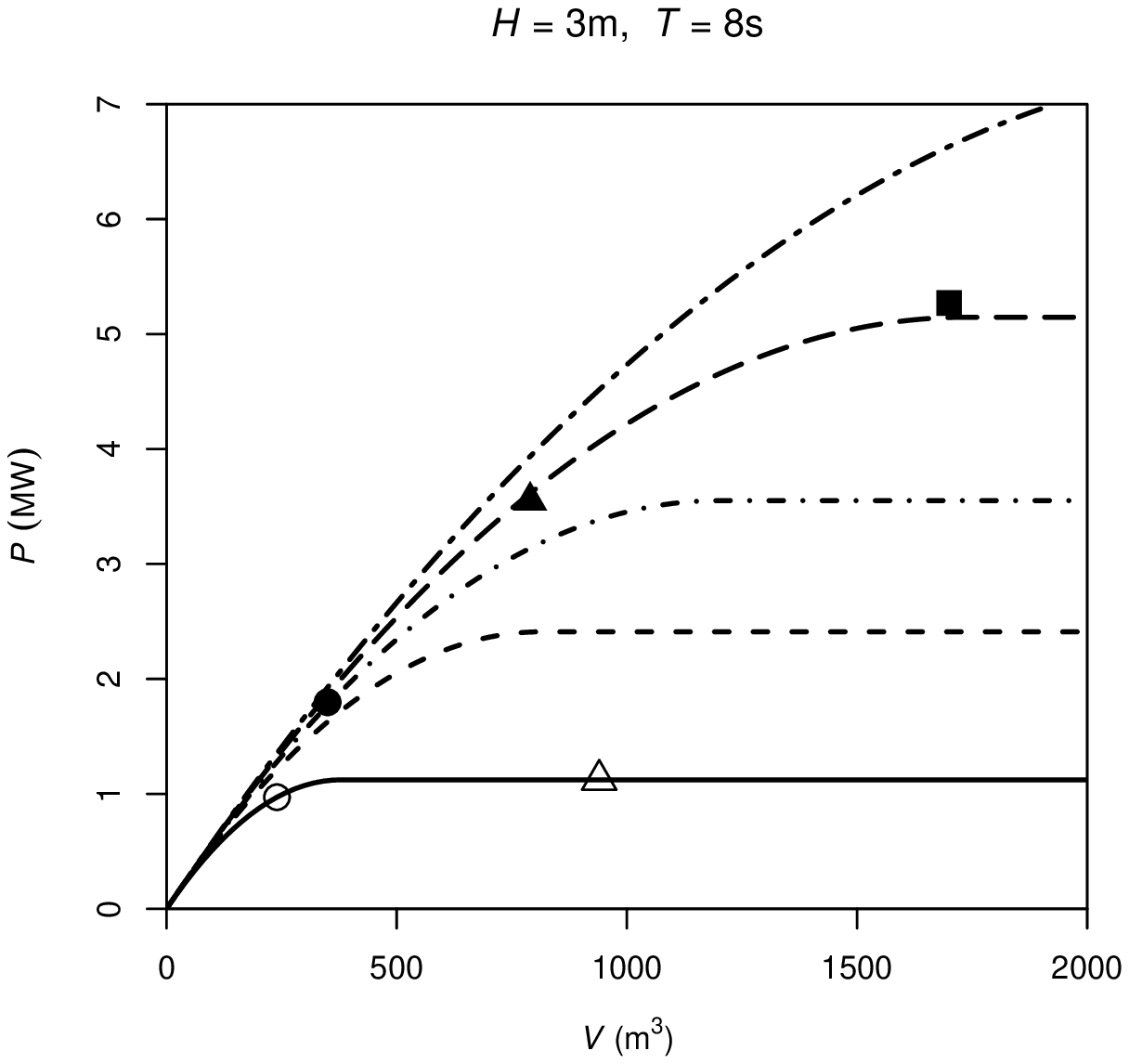}\includegraphics[width=0.5\textwidth]{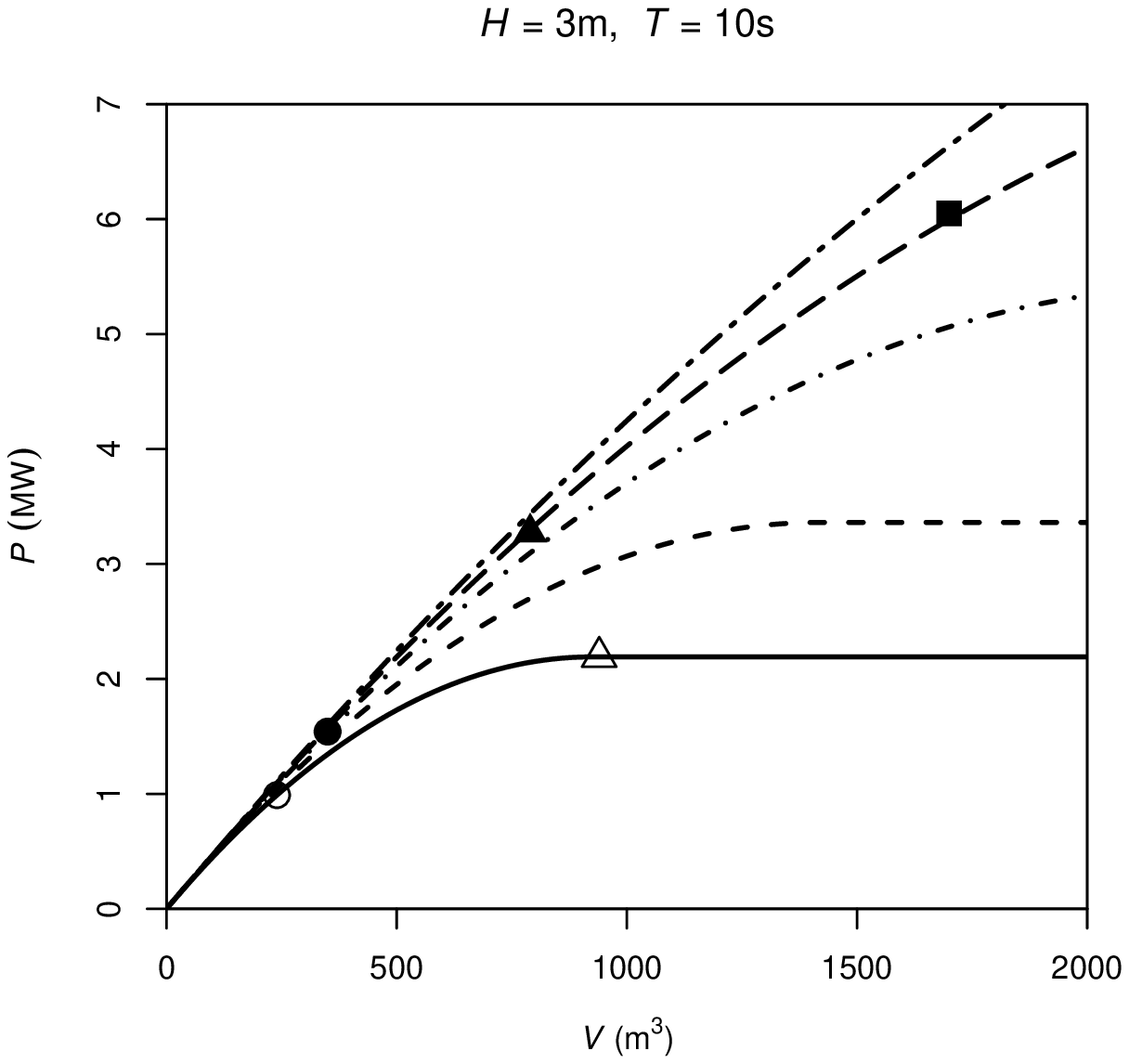}
\par\end{centering}

\caption{\label{fig:P_V_devs}Absorbed power, $P$, from (\ref{eq:w_l_V^*})
as a function of half-swept volume, $V$, for WECs with sizes described
in \T\ref{tab:devices} operating in four different $(H,T)$ incident
waves.}
\ifthenelse{\boolean{includeNotes}}{

\textcolor{green}{Figures produced by \textasciitilde{}/opd/documents/line\_absorbers/R-analysis/power\_plots\_AOR.R.}

}

\end{figure*}

Figure~\ref{fig:w^*_V^*_devs} shows the points of maximum dimensionless
capture width plotted against dimensionless volume for the sizes of
WECs given in \T\ref{tab:devices} operating in the four incident
waves with $(H,T)$ equal to $(2\mbox{m},8\mbox{s})$, $(2\mbox{m},10\mbox{s})$,
$(3\mbox{m},8\mbox{s})$ and $(3\mbox{m},10\mbox{s})$. These four
incident waves broadly correspond to those that would occur close
to the centres of the available energy resource when weighted by annual
average occurrence. They have incident power in ratios of $1,$ $1.25$,
$2.25$ and $2.81$ respectively. Also, shown in \F\ref{fig:w^*_V^*_devs}
are contours of constant dimensionless length. The set of four plots
show how the dimensionless volume and maximum capture width of a particular
WEC depends on the wave height and period of the incident wave. The
plots show that the largest volume point absorber appears below the
line $w^{*}=V^{*}$ and is, therefore, operating in its unlimited
volume regime and unnecessarily large for all but the $(3\mbox{m},10\mbox{s})$
incident wave. The largest volume line-absorber, however, is only
unnecessarily large for the $(2\mbox{m},8\mbox{s})$ incident wave,
even though it is nearly twice the volume of the largest point absorber.
The smallest point absorber has its maximum capture width limited
by its small volume in all but the $(2\mbox{m},8\mbox{s})$ incident
wave. The mid-sized line absorber has less volume than the large point
absorber, yet its maximum capture width is about four times larger
in the $(2\mbox{m},8\mbox{s})$ wave, three times larger in the $(3\mbox{m},8\mbox{s})$
wave, twice as large in the $(2\mbox{m},10\mbox{s})$ wave, and one
and half times as large in the $(3\mbox{m},10\mbox{s})$ wave. For
further clarity, \F\ref{fig:P_V_devs} shows points of dimensional
maximum absorbed power plotted against dimensional volume for the
size of WECs given in \T\ref{tab:devices} operating in the four
incident waves as used in \F\ref{fig:w^*_V^*_devs}. A set of conveniently
spaced contours of constant length are also shown in these plots.
This set of four plots shows how the maximum absorbed power of a particular
size of WEC depends on the wave height and period of the incident
wave.

\section{Conclusions }

The theoretical maximum absorbed power and capture width of a limited
volume attenuating line absorber heaving in a travelling wave mode
in the presence of progressive regular incident wave has been derived
in the frequency domain using the linearised theory of the interaction
of water waves and structures. The results are presented in dimensionless
form which has the advantage of reducing the number of dependent variables
from four to just two: dimensionless length, $l^{*}$, and dimensionless
half-swept volume, $V^{*}$. In the zero-length limit the results
for the limited volume line absorber reduce to those for a point absorber;
and in the zero-volume limit they reduce to the result expressed by
Budal's upper bound. It is shown that, in dimensionless units, the
maximum dimensionless capture width of a point absorber is $w^{*}=1$
and that the smallest volume required to achieve this capture width
is $V^{*}=1$. The dimensionless capture width and volume both depend
on the incident wave: a value of $w^{*}=1$ corresponds to a width
of about $16\%$ of the wavelength; a value of $V^{*}=1$ corresponds
to about $2.5\%$ of the rectangular volume given by the square of
the wavelength times the wave amplitude. Increasing the volume of
a point absorber beyond $V^{*}=1$ gives no increase in capture width.
For an attenuating line absorber, however, in the limit of infinite
length the maximum capture width is $w^{*}=2V^{*}$. Thus, there is
no limit to the capture width of a line absorber provided it has sufficient
volume and length. It is shown that a particular fixed-length line
absorber will have a maximum capture width of $w^{*}>1$ and that
the smallest volume required to achieve this capture width is $V^{*}=w^{*}$
.

This has profound implications for the economics of power generation
from wave energy converters. Even though small volume point absorbers
are more efficient than larger volume point absorbers in terms of
power absorbed per unit swept-volume of device, engineering limitations
make installation and operation of very large numbers of very small
devices excessively costly and uneconomic. A point absorber has a
limit on its capture width and usable volume, whereas, in theory,
line absorbers can be indefinitely scaled-up in volume and length
to give unlimited capture widths. This theoretical advantage of line
absorbers over point absorbers applies to realistically sized wave
energy converters currently in operation. Thus, line absorbers can
be progressively scaled-up in volume and length to give increasingly
large capture widths while retaining dimensions compatible with cost-effective
engineering. For example, a line absorber with a volume equal to the
maximum usable volume of a point absorber and a length equal to twice
the wavelength of the incident wave can absorb nearly $80\%$ more
power than a point absorber of the same volume. Doubling this line
absorber's volume gives over three times the absorbed power of the
point absorber, and tripling it gives over four times the absorbed
power. Thus, it is clear that by installing fewer larger line absorbers
valuable economies of scale can be achieved from line absorbers that
are not achievable from point absorbers.

\appendix

\section{Integrals by the method of stationary phase \label{app:stat_phase_and_I}}

The terms on the right-hand side of (\ref{eq:phi_d_phi}) are evaluated
in cylindrical polar coordinates as follows. Substituting $x=r\cos\theta$
into (\ref{eq:varphi_0}), $\varphi_{0}$ is written as \begin{equation}
\varphi_{0}=-i\frac{g}{\omega}\frac{\cosh(k(z+h))}{\cosh(kh)}\mathrm{e}^{ikr\cos\theta}.\label{eq:varphi_0^*_r}\end{equation}
Also, using (\ref{eq:varphi_0}) along with $n_{x}=\cos\theta$ and
$n_{z}=0$ on the control surface $S_{C}$, and the observation that
$\bar{\varphi}_{0}$ is not a function of $y$, it follows that\begin{multline}
\frac{\partial\bar{\varphi}_{0}}{\partial n}=\frac{gk}{\omega}\frac{\cosh(k(z+h))}{\cosh(kh)}\,\cos\theta\,\mathrm{e}^{-ikr\cos\theta},\\
\quad\mbox{on }S_{C}.\label{eq:grad_varphi_0^*_r}\end{multline}

Next, using\[
\mathbf{n}=\hat{\mathbf{e}}_{r}\quad\mbox{on }S_{C},\]
and the gradient operator in cylindrical coordinates\[
\boldsymbol{\nabla}=\hat{\mathbf{e}}_{r}\frac{\partial}{\partial r}+\hat{\mathbf{e}}_{\theta}\frac{1}{r}\frac{\partial}{\partial\theta}+\hat{\mathbf{e}}_{z}\frac{\partial}{\partial z},\]
it follows that\[
\mathbf{n}\cdot\boldsymbol{\nabla}\bar{\varphi}_{z}=\frac{\partial\bar{\varphi}_{z}}{\partial n}\equiv\frac{\partial\bar{\varphi}_{z}}{\partial r}\quad\mbox{on }S_{C}.\]
Evaluating $\partial\bar{\varphi}_{z}/\partial n$ by differentiating
(\ref{eq:phi_alpha_bess}) with respect to $r$, it can be shown that\begin{alignat}{1}
\frac{\partial\bar{\varphi}_{z}}{\partial n} & \sim\frac{igk^{3}bl}{\omega\left(2\pi kr\right)^{1/2}}\frac{\cosh(k(z+h))}{\cosh(kh)}\times\nonumber \\
 & j_{0}\negmedspace\left(\frac{kl}{2}\left(1-\cos\theta\right)\right)\,\mathrm{e}^{-ikr+i\pi/4}\quad\mbox{on }S_{C},\label{eq:grad_varphi_z^*}\end{alignat}
where $j_{0}$ is the zeroth spherical Bessel function of the first
kind and, since $kr\to\infty$ on $S_{C}$, only the highest order
terms in $kr$ have been retained.

Combining (\ref{eq:phi_alpha_bess}), (\ref{eq:varphi_0^*_r}), (\ref{eq:grad_varphi_0^*_r})
and (\ref{eq:grad_varphi_z^*}), the three integrands on the right-hand
side of (\ref{eq:phi_d_phi}) can now be simplified and written as\begin{alignat}{1}
\varphi_{0}\,\frac{\partial\bar{\varphi}_{z}}{\partial n} & \sim\frac{g^{2}k^{3}bl}{\omega^{2}\left(2\pi kr\right)^{1/2}}\left(\frac{\cosh(k(z+h))}{\cosh(kh)}\right)^{2}\times\nonumber \\
 & j_{0}\negmedspace\left(\frac{kl}{2}\left(1-\cos\theta\right)\right)\,\mathrm{e}^{-ikr\left(1-\cos\theta\right)+i\pi/4},\label{eq:int_b1}\\
\varphi_{z}\,\frac{\partial\bar{\varphi}_{0}}{\partial n} & \sim-\frac{g^{2}k^{3}bl}{\omega^{2}\left(2\pi kr\right)^{1/2}}\left(\frac{\cosh(k(z+h))}{\cosh(kh)}\right)^{2}\times\nonumber \\
 & \cos\theta\, j_{0}\negmedspace\left(\frac{kl}{2}\left(1-\cos\theta\right)\right)\,\mathrm{e}^{ikr\left(1-\cos\theta\right)-i\pi/4},\label{eq:int_b2}\\
\varphi_{z}\,\frac{\partial\bar{\varphi}_{z}}{\partial n} & \sim-\frac{ik}{2\pi kr}\left(\frac{gk^{2}bl}{\omega}\,\frac{\cosh(k(z+h))}{\cosh(kh)}\right)^{2}\times\nonumber \\
 & j_{0}^{2}\negmedspace\left(\frac{kl}{2}\left(1-\cos\theta\right)\right).\label{eq:int_b3}\end{alignat}

In cylindrical polar coordinates the integral over $S_{C}$ on the
left-hand of (\ref{eq:phi_d_phi}) is expressed as\begin{equation}
\iint_{S_{C}}\phi\,\frac{\partial\bar{\phi}}{\partial n}\, dS=\int_{-h}^{0}\int_{-\pi}^{\pi}\phi\,\frac{\partial\bar{\phi}}{\partial n}\, r\, d\theta\, dz.\label{eq:gen_int_polar_coords}\end{equation}
Each of (\ref{eq:int_b1})--(\ref{eq:int_b3}) has the same $z$-dependency.
Integrating out this $z$-dependency gives, after simplification and
substitution of the dispersion relationship from (\ref{eq:dispersion}),
\begin{equation}
\int_{-h}^{0}\frac{\cosh^{2}(k(z+h))}{\cosh^{2}(kh)}\, dz=\frac{\omega c_{g}}{gk},\label{eq:int_z}\end{equation}
where $c_{g}$ is the group velocity given by (\ref{eq:c_g}). The
integrals over $\theta$ of (\ref{eq:int_b1})--(\ref{eq:int_b3})
are evaluated using the method of stationary phase, which states that
for real-valued smooth functions $f$ and $g$, when $f'(\theta_{0})=0$
and $K\to\infty$,\begin{multline}
\int_{-\infty}^{\infty}g(\theta)\,\mathrm{e}^{iKf(\theta)}\, d\theta\\
\sim g(\theta_{0})\sqrt{\frac{2\pi}{Kf''(\theta_{0})}}\mathrm{e}^{i\left(Kf(\theta_{0})+\pi/4\right)},\\
\mbox{with }f'(\theta_{0})=0,\, K\to\infty.\label{eq:stat_phase_gen}\end{multline}
 Substituting $K=kr$, $g(x)=j_{0}\negmedspace\left(\frac{kl}{2}\left(1-\cos\theta\right)\right)$,
$f(x)=\cos\theta-1$ and $\theta_{0}=0$ into (\ref{eq:stat_phase_gen})
gives the integral of (\ref{eq:int_b1}) over $\theta$ as\begin{multline}
\int_{-\pi}^{\pi}j_{0}\negmedspace\left(\frac{kl}{2}\left(1-\cos\theta\right)\right)\,\mathrm{e}^{-ikr(1-\cos\theta)}\, d\theta\\
\sim\sqrt{\frac{2\pi}{kr}}\mathrm{e}^{-i\pi/4}.\label{eq:int_j0_1}\end{multline}
  Substituting $K=kr$, $g(x)=\cos\theta\, j_{0}\negmedspace\left(\frac{kl}{2}\left(1-\cos\theta\right)\right)$,
$f(x)=1-\cos\theta$, and $\theta_{0}=0$ into (\ref{eq:stat_phase_gen})
gives the integral over $\theta$ of (\ref{eq:int_b2}) as\begin{multline}
\int_{-\pi}^{\pi}\cos\theta\, j_{0}\negmedspace\left(\frac{kl}{2}\left(1-\cos\theta\right)\right)\mathrm{e}^{ikr(1-\cos\theta)}\, d\theta\\
\sim\sqrt{\frac{2\pi}{kr}}\mathrm{e}^{i\pi/4}.\label{eq:int_j0_2}\end{multline}
For convenience of notation the integral of (\ref{eq:int_b3}) over
$\theta$ is denoted by $I(kl)$ as defined by the integral in (\ref{eq:def_I_kl}).
As stated by \citeauthor{Farley82} \citep[\Eq(20)]{Farley82}, $I(kl)$
can be expressed analytically as in (\ref{eq:I_kl_analytic}) in terms
of first-order Bessel functions of the first kind, $J_{0}$ and $J_{1}$.
 Since the $kl\to0$ limits of the Bessel functions are \begin{eqnarray*}
\lim_{kl\to0}J_{0}(kl) & = & 1,\\
\lim_{kl\to0}\frac{J_{1}(kl)}{kl} & = & \frac{1}{2},\end{eqnarray*}
the $kl\to0$ limit of (\ref{eq:I_kl_analytic}) is \begin{equation}
\lim_{kl\to0}I(kl)\to1.\label{eq:I_limit}\end{equation}

Equations~(\ref{eq:int_z}), (\ref{eq:int_j0_1}), (\ref{eq:int_j0_2})
and (\ref{eq:def_I_kl}) are now used, along with (\ref{eq:gen_int_polar_coords}),
to evaluate the integrals of (\ref{eq:int_b1})--(\ref{eq:int_b3})
over $S_{C}$. The results are\begin{alignat*}{1}
\int_{-h}^{0}\int_{-\pi}^{\pi}\varphi_{0}\,\frac{\partial\bar{\varphi}_{z}}{\partial n}\, r\, d\theta\, dz & \sim\frac{c_{g}gkbl}{\omega},\\
\int_{-h}^{0}\int_{-\pi}^{\pi}\varphi_{z}\,\frac{\partial\bar{\varphi}_{0}}{\partial n}\, r\, d\theta\, dz & \sim-\frac{c_{g}gkbl}{\omega},\\
\int_{-h}^{0}\int_{-\pi}^{\pi}\varphi_{z}\,\frac{\partial\bar{\varphi}_{z}}{\partial n}\, r\, d\theta\, dz & \sim-\frac{ic_{g}gk^{3}b^{2}l^{2}}{\omega}\, I(kl).\end{alignat*}
Substituting these expressions into (\ref{eq:phi_d_phi}) and the
result into (\ref{eq:dE/dt_Sc_2}) gives (\ref{eq:dE/dt_lineAten_0}).

\biboptions{sort&compress}
\bibliographystyle{elsarticle-num-names}
\bibliography{line_absorbers_AOR}

\begin{thebibliography}{15}
\providecommand{\natexlab}[1]{#1}
\providecommand{\url}[1]{\texttt{#1}}
\providecommand{\urlprefix}{URL }
\expandafter\ifx\csname urlstyle\endcsname\relax
  \providecommand{\doi}[1]{doi:\discretionary{}{}{}#1}\else
  \providecommand{\doi}[1]{doi:\discretionary{}{}{}\begingroup
  \urlstyle{rm}\url{#1}\endgroup}\fi
\providecommand{\bibinfo}[2]{#2}

\bibitem[{Evans(1976)}]{Evans1976}
\bibinfo{author}{D.~V. Evans}, \bibinfo{title}{A theory for wave-power
  absorption by oscillating bodies}, \bibinfo{journal}{Journal of Fluid
  Mechanics} \bibinfo{volume}{77} (\bibinfo{year}{1976})
  \bibinfo{pages}{1--25}.

\bibitem[{Mei(1976)}]{Mei1976}
\bibinfo{author}{C.~C. Mei}, \bibinfo{title}{Power extraction from water
  waves}, \bibinfo{journal}{Journal of Ship Research}
  \bibinfo{volume}{20}~(\bibinfo{number}{2}) (\bibinfo{year}{1976})
  \bibinfo{pages}{63--66}.

\bibitem[{Newman(1976)}]{Newman1976}
\bibinfo{author}{J.~N. Newman}, \bibinfo{title}{The Interaction of Stationary
  Vessels with Regular Waves}, in: \bibinfo{booktitle}{Proceedings of the 11th
  Symposium on Naval Hydrodynamics}, \bibinfo{publisher}{Mechanical Engineering
  Publications Limited}, \bibinfo{address}{London, UK},
  \bibinfo{pages}{491--501}, \bibinfo{year}{1976}.

\bibitem[{Evans(1981)}]{Evans1981}
\bibinfo{author}{D.~V. Evans}, \bibinfo{title}{Maximum wave-power absorption
  under motion constraints}, \bibinfo{journal}{Applied Ocean Research}
  \bibinfo{volume}{3} (\bibinfo{year}{1981}) \bibinfo{pages}{200}.

\bibitem[{Pizer(1993)}]{Pizer93}
\bibinfo{author}{D.~J. Pizer}, \bibinfo{title}{Maximum wave-power absorption of
  point absorbers under motion constraints}, \bibinfo{journal}{Applied Ocean
  Research} \bibinfo{volume}{15} (\bibinfo{year}{1993})
  \bibinfo{pages}{227--234}.

\bibitem[{Farley(1982)}]{Farley82}
\bibinfo{author}{F.~J.~M. Farley}, \bibinfo{title}{Wave energy conversion by
  flexible resonant rafts}, \bibinfo{journal}{Applied Ocean Research}
  \bibinfo{volume}{4}~(\bibinfo{number}{1}) (\bibinfo{year}{1982})
  \bibinfo{pages}{57--63}.

\bibitem[{Newman(1979)}]{Newman1979}
\bibinfo{author}{J.~N. Newman}, \bibinfo{title}{Absorption of wave energy by
  elongated bodies}, \bibinfo{journal}{Applied Ocean Research}
  \bibinfo{volume}{4} (\bibinfo{year}{1979}) \bibinfo{pages}{189--196}.

\bibitem[{Rainey(2001)}]{Rainey01}
\bibinfo{author}{R.~C.~T. Rainey}, \bibinfo{title}{The Pelamis wave energy
  converter: it may be jolly good in practice, but will it work in theory?},
  in: \bibinfo{booktitle}{Proceedings of the 16th International Workshop on
  Water Waves and Floating Bodies}, \bibinfo{address}{Hiroshima, Japan},
  \bibinfo{pages}{1--6}, \bibinfo{year}{2001}.

\bibitem[{Mei(1989)}]{Mei89}
\bibinfo{author}{C.~C. Mei}, \bibinfo{title}{The applied dynamics of ocean
  surface waves}, \bibinfo{publisher}{Wiley-Interscience, John Wiley \& Sons,
  Inc.}, \bibinfo{note}{iSBN 0-471-06407-6}, \bibinfo{year}{1989}.

\bibitem[{McIver(1994)}]{McIver1994}
\bibinfo{author}{P.~McIver}, \bibinfo{title}{Low-frequency asymptotics of
  hydrodynamic forces on fixed and floating structures}, in:
  \bibinfo{editor}{M.~Rahman} (Ed.), \bibinfo{booktitle}{Waves Engineering},
  \bibinfo{publisher}{Computational Mechanics Publications},
  \bibinfo{pages}{1--49}, \bibinfo{year}{1994}.

\bibitem[{Falnes(1993)}]{Falnes1993}
\bibinfo{author}{J.~Falnes}, \bibinfo{title}{Small is beautiful: How to make
  wave energy economic}, in: \bibinfo{booktitle}{European wave energy
  symposium}, \bibinfo{address}{Edinburgh, Scotland},
  \bibinfo{pages}{367--372}, \bibinfo{year}{1993}.

\bibitem[{Inc.(2011)}]{pb150}
\bibinfo{author}{O.~P.~T. Inc.},
  \urlprefix\url{http://www.oceanpowertechnologies.com/pb150.htm},
  \bibinfo{note}{a 50\% submerged torus of about diameter 11m and height 2.5m},
  \bibinfo{year}{2011}.

\bibitem[{Ltd.(2011)}]{wavebob}
\bibinfo{author}{W.~Ltd.}, \urlprefix\url{http://wavebob.com/key-features/},
  \bibinfo{note}{a 50\% submerged torus of about diameter 20m and height 8m},
  \bibinfo{year}{2011}.

\bibitem[{{Pelamis~Wave~Power~Ltd.}(2011{\natexlab{a}})}]{pelamisFSP}
\bibinfo{author}{{Pelamis~Wave~Power~Ltd.}},
  \urlprefix\url{http://www.pelamiswave.com/our-technology/development-history%
}, \bibinfo{note}{a 30\% submerged line absorber of length 120m and diameter
  3.5m}, \bibinfo{year}{2011}{\natexlab{a}}.

\bibitem[{{Pelamis~Wave~Power~Ltd.}(2011{\natexlab{b}})}]{pelamisP2}
\bibinfo{author}{{Pelamis~Wave~Power~Ltd.}},
  \urlprefix\url{http://www.pelamiswave.com/our-technology/the-p2-pelamis},
  \bibinfo{note}{a 35\% submerged line absorber of length 180m and diameter
  4m}, \bibinfo{year}{2011}{\natexlab{b}}.

\end{thebibliography}

\end{document}